\documentclass[twocolumn,english,aps,prb,reprint, superscriptaddress,showpacs,longbibliography,showkeys]{revtex4-2}
\usepackage{amsmath,amssymb,bbm,mathrsfs,bm,braket,color,graphicx,comment,xcolor,dsfont,multirow}
\usepackage[colorlinks,citecolor=blue,urlcolor=blue,linkcolor = blue]{hyperref}
\usepackage{enumitem}
\usepackage[mathscr]{euscript}
\usepackage[T1]{fontenc}
\usepackage{siunitx}
\usepackage{placeins}

\begin{document}

\title{Scalable Parity Architecture With a Shuttling-Based Spin Qubit Processor}

\author{Florian Ginzel}
    \affiliation{Parity Quantum Computing Germany GmbH, Schauenburgerstraße 6, 20095 Hamburg, Germany}

\author{Michael Fellner}
    \affiliation{Parity Quantum Computing GmbH, Rennweg 1, Top 314, 6020 Innsbruck, Austria}
    \affiliation{Institute for Theoretical Physics, University of Innsbruck, 6020 Innsbruck, Austria}

\author{Christian Ertler}
    \affiliation{Parity Quantum Computing Germany GmbH, Schauenburgerstraße 6, 20095 Hamburg, Germany}

\author{Lars R. Schreiber}
    \affiliation{JARA-FIT Institute for Quantum Information, Forschungszentrum J\"ulich GmbH and RWTH Aachen University, Aachen, Germany}
    \affiliation{ARQUE Systems GmbH, 52074 Aachen, Germany}

\author{Hendrik Bluhm}
    \affiliation{JARA-FIT Institute for Quantum Information, Forschungszentrum J\"ulich GmbH and RWTH Aachen University, Aachen, Germany}
    \affiliation{ARQUE Systems GmbH, 52074 Aachen, Germany}

\author{Wolfgang Lechner}
    \affiliation{Parity Quantum Computing Germany GmbH, Schauenburgerstraße 6, 20095 Hamburg, Germany}
    \affiliation{Parity Quantum Computing GmbH, Rennweg 1, Top 314, 6020 Innsbruck, Austria}
    \affiliation{Institute for Theoretical Physics, University of Innsbruck, 6020 Innsbruck, Austria}

\date{\today}

\begin{abstract}
    Motivated by the prospect of a two-dimensional square-lattice geometry for semiconductor spin qubits, we explore the realization of the Parity Architecture with quantum dots (QDs). %This is part of the endeavor of developing architectures that advance the utilization of spin qubits for quantum computing while harnessing their advantages, such as their fast timescales---especially of the nearest-neighbor interaction---and small size.
    We present sequences of spin shuttling and quantum gates that implement the Parity Quantum Approximate Optimization Algorithm (QAOA) on a lattice constructed of identical unit cells, such that the circuit depth is always constant. We further develop a detailed error model for a hardware-specific analysis of the Parity Architecture and we estimate the errors during one round of Parity QAOA. The model includes a general description of the shuttling errors as a function of the probability distribution function of the valley splitting, which is the main limitation for the performance. We compare our approach to a superconducting transmon qubit chip and we find that with high-fidelity spin shuttling the performance of the spin qubits is competitive or even exceeds the results of the transmons. %We find that a physical single-qubit error probability of $\lesssim 5\%$ can be expected already in near term-devices, mainly limited by the low valley splitting. 
    Finally, we discuss the possibility of decoding the logical quantum state and of quantum error mitigation. We find that already with near-term spin qubit devices a sufficiently low physical error probability can be expected to reliably perform Parity QAOA with a short depth in a regime where the success probability compares favorably to standard QAOA.
\end{abstract}

\maketitle

\section{Introduction}

In the pursuit of quantum computing, the question of the ideal hardware platform is still unanswered. Among the contestants, spin qubits in gate-defined quantum dots (QDs)~\cite{RevModPhys.95.025003} stand out with their unique promise of leveraging the sophisticated manufacturing capabilities of the semiconductor industry once a design based on scalable building blocks is devised~\cite{Vandersypen2017,Crawford2023,ArqueProposal}. This, combined with their small size of only a few tens of nanometers per QD, may allow for the fabrication of quantum computers with millions of qubits that can easily be mass-produced~\cite{Zwerver2022}.

While the gate fidelities are approaching the threshold to quantum error correction~\cite{Xue2022,Mills2022,Noiri2022}, serious challenges must still be overcome on the road to spin-based quantum computing. The most prominent are environmental electric noise~\cite{PhysRevB.100.165305,Connors2022}, cross-talk and residual exchange interaction within dense arrays of qubits~\cite{Huang2019,PhysRevB.104.045420,PhysRevB.105.085414,Heinz2023}, and the demanding space requirements of the voltage gates and control electronics~\cite{Vandersypen2017,Crawford2023,ArqueProposal}.

A possible means of addressing those problems could be spin shuttling, coherently moving the qubits between sites --- with different functionality --- on the chip on demand. Shuttling can be realized either in the conveyor-mode where a sliding potential well smoothly displaces the qubit~\cite{Taylor2005,Seidler2022}, or as a bucket brigade by coherent tunneling between adjacent QDs~\cite{Flentje2017,Fujita2017,Mills2019}. While the latter variant requires a high degree of individual control~\cite{PhysRevB.102.195418,PRXQuantum.4.020305}, conveyor-mode shuttling with potentials formed by dedicated gates is showing promise for success~\cite{Seidler2022,Struck2023,Xue2023}. In silicon heterostructures, the degenerate conduction band minima lead to an additional pseudospin, namely the valley degree of freedom, whose splitting is determined by the microscopic properties of the interface~\cite{PhysRevLett.88.027903,PhysRevB.81.115324}. Local minima of the valley splitting can be a major challenge for conveyor-mode shuttling, however, their occurrence can be reduced by by engineering the semiconductor heterostructure~\cite{PhysRevB.80.081305,PaqueletWuetz2022,Lima_2023_1,Lima_2023_2,PhysRevB.104.085406,McJunkin2022} or adjusting shuttling trajectories~\cite{vollmer2024}.

\begin{figure}
    \centering
    \includegraphics[width=0.48\textwidth]{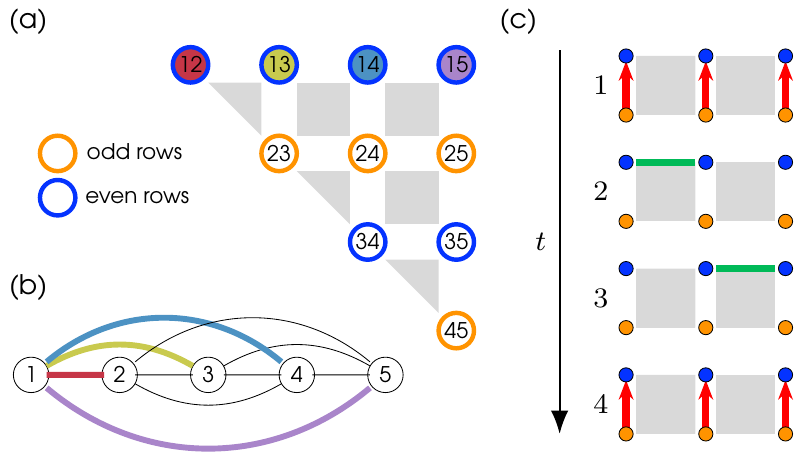}
    \caption{(a) Visualization of the Parity Architecture. (b) An optimization problem on logical qubits 1-5 can be expressed by assigning one physical qubit $ij$ in panel (a) to each logical interaction $J_{ij}$. This allows implementing the problem Hamiltonian with local fields, while constraints must be enforced on plaquettes of four or three adjacent qubits (gray). Note that it is also possible to compile higher order interaction terms to an arbitrary chip layout~\cite{ender_parity_2023}. For the decomposition of the constraints in Parity QAOA the qubits are organized in rows 0, 1, 2, ... (blue and orange).
    (c) Half of the circuit that enforces the constraints of the parity transformation during Parity QAOA~\cite{unger_low-depth_2022}. The lattice is separated into ribbons of adjacent rows on which CNOT (red arrows pointing from control to target) and ZZ gates (green lines) are applied in successive steps 1-4. All ribbons with even-numbered top row can be treated in parallel, followed by all ribbons with an odd-numbered top row in parallel.}
    \label{fig_LHZ}
\end{figure}

A major objective in the development of spin qubits is the creation of connectivity between qubits in two dimensions. Inevitably for error-corrected quantum computing~\cite{Campbell2017}, this milestone could also make the Parity Architecture~\cite{lechner_quantum_2015, ender_parity_2023} a viable way to advance the performance of spin qubits. In the Parity Architecture the logical state of the quantum computer is encoded by physical qubits that represent the parity of the logical spins [see Fig.~\ref{fig_LHZ}(a)-(b)]. While introducing a qubit overhead, this encoding removes the requirement for long-distance interactions and the redundant information allows for quantum error mitigation~\cite{PhysRevA.108.032408} and quantum error correction~\cite{PhysRevLett.129.180503} for bit-flip errors. Furthermore, it allows the execution of the Parity Quantum Approximate Optimization Algorithm (QAOA)~\cite{lechner_quantum_2020,unger_low-depth_2022} and reduces the circuit depth of cornerstone algorithms such as the quantum Fourier transform~\cite{PhysRevA.106.042442,Messinger2023}.

The QAOA is a gate-based algorithm for solving combinatorial optimization problems on a digital quantum computer~\cite{farhi2014quantum,farhi2016quantum,a12020034}, inspired by adiabatic quantum computing, where a quantum state is evolved adiabatically under a Hamiltonian representing the cost function of the optimization problem in order to approximate its ground state~\cite{doi:10.1126/science.1057726}. Here, adiabatic time evolution is replaced by an alternating sequence of parameterized small angle rotations corresponding to a problem and driver Hamiltonian respectively. The parameters are then optimized in a quantum-classical feedback loop. The first proof-of-principle demonstrations of QAOA were shown on existing quantum hardware~\cite{doi:10.1073/pnas.2006373117,Harrigan2021,doi:10.1126/science.abo6587} and the algorithm may be a suitable candidate to prove a quantum advantage for non-trivial problems with a few hundred qubits and gate fidelity below the error correction threshold~\cite{farhi2016quantum,Preskill2018quantumcomputingin,Weidenfeller2022scalingofquantum}. However, it remains challenging to achieve the connectivity required for an arbitrary problem Hamiltonian, which contributes to a non-optimal scaling of resources with the problem size and complexity~\cite{Harrigan2021,Weidenfeller2022scalingofquantum}. This problem is alleviated by the Parity Architecture~\cite{lechner_quantum_2015,lechner_quantum_2020}.

We find that spin qubits can efficiently implement the Parity Architecture even compared to more mature platforms since their native gates naturally fit the demands for Parity QAOA and thus require little additional transpilation. Different strategies are currently under investigation for realizing a two-dimensional lattice of spin qubits, which perfectly suits the Parity Architecture. Its possibility of working with local fields and nearest neighbor interactions allows leveraging the fast, albeit short-range, two-qubit gates, one of the main advantages of spin qubits. 

In this article, we investigate the performance of Parity QAOA on two scalable spin qubit architectures based on shuttling and modularization of the chip. We develop the shuttling sequences for implementing the algorithm on these architectures and we show that Parity QAOA can be efficiently executed even though the topology of the chips is not a square lattice, as required in the original proposals. This result is of relevance for many hardware platforms that suffer from a low connectivity, and it is of particular interest for spin qubits, where the realization of fully connected two-dimensional arrays is hindered by cross-talk and the space required for the fan-out of the voltage gates. 

The performance of Parity QAOA has been analyzed generic error models that do not allow to reliably gauge the performance of real hardware. Here, instead, we introduce a realistic error model based on recent experimental, theoretical and computational results for an in-depth performance analysis of Parity QAOA. Based on our error model the error probability of a single physical qubit is estimated. We find that with just slightly optimistic assumptions for the future development of the spin qubit coherence time, a full round of Parity QAOA is feasible on both architectures, with the possibility to further enhance the performance by error mitigation. We note that the error probability is both in a regime where simulations estimate a higher success rate for Parity QAOA compared to standard QAOA~\cite{PhysRevA.108.032408} and where Parity QAOA can address non-trivial problems~\cite{StilckFranca2021}.

Although the main result of the article -- the analysis of the algorithmic performance -- is specific to the hardware under consideration, this prediction of good performance of an algorithm tailored to the Parity Architecture under a realistic noise model suggests that the Parity Architecture is a suitable candidate for demonstrations on different types of noisy near-term quantum hardware. In particular, we make a comparison with a chip consisting of capacitively coupled transmons in the same layout as the modular spin qubit architecture. For an optimal shuttling velocity and an engineered valley splitting, the performance of Parity QAOA with spin and superconducting qubits are comparable and a shuttling-based spin qubit processor can surpass even optimistic assumptions for the transmon qubits.

The remainder of this article is organized as follows. In Sec.~\ref{sec_intro_parity} a brief introduction to Parity quantum computing is provided, followed by our proposed implementation of the Parity QAOA algorithm on a spin qubit quantum processor in Sec.~\ref{sec_architectures}. In particular, Sec.~\ref{sec_spin_bus} focuses on the case of a sparse shuttling-based architecture and Sec.~\ref{sec_modular} focuses on the case of a modular architecture where the registers are connected by spin shuttling. In Sec.~\ref{sec_error_model} the error model for all relevant processes is introduced. In Sec.~\ref{sec_error_evaluation} the performance of the QAOA on both architectures is investigated in the presence of realistic errors and in Sec.~\ref{sec_transmons} the results are put into context with superconducting transmon qubits . In Sec.~\ref{sec_decoding}, the possibilities for error mitigation and the decoding of the quantum state are discussed. Finally, in Sec.~\ref{sec_conclusions}, the results are summarized and the paper is concluded.

\section{Introduction to Parity Quantum Computing\label{sec_intro_parity}}

Here we provide an introduction to the Parity Architecture in general (Sec.~\ref{sec_parity_general}) and the Parity QAOA in particular (Sec.~\ref{sec_parity_QAOA}).

\subsection{The Parity Architecture\label{sec_parity_general}}

Consider an optimization problem encoded in a spin-glass problem Hamiltonian with $N$ qubits and $K$ interactions of the form
\begin{eqnarray}
    H_\mathrm{op} = \,&& \sum_{i=1}^N J_i \sigma_z^{(i)} +\sum_{j<i} J_{ij} \sigma_z^{(i)} \sigma_z^{(j)} \nonumber\\
    && + \sum_{k<j<i} J_{ijk} \sigma_z^{(i)} \sigma_z^{(j)} \sigma_z^{(k)}+\dots,
\end{eqnarray}
where the coefficients $J_i, J_{ij}, \dots$ denote the coupling strengths and $\sigma_z^{(i)}$ denotes the Pauli-Z operator acting on logical qubit $i$.
The Parity Architecture~\cite{lechner_quantum_2015, ender_parity_2023} maps this Hamiltonian to the parity Hamiltonian
\begin{equation}
    H_\text{parity} = \sum_{m=1}^K \tilde J_m \tilde \sigma_z^{(m)} + c \sum_{l=1}^{K-N+D}  C_l
\end{equation}
with $K$ physical qubits, where each interaction between logical qubits in $H_\mathrm{op}$ was mapped to a local field term with strength $\tilde J_m$ on physical qubit $m$ with associated Pauli-Z operator $\tilde \sigma_z^{(m)}$, e.g., ${J_{ij}\sigma_z^{(i)}\sigma_z^{(j)} \mapsto J_m \tilde\sigma_z^{(m)}}$, and ${c>0}$ denotes a constant. The physical qubits represent the parity of a set of logical qubits, i.e. their $\sigma_z$-eigenvalue indicates if the respective logical qubits are in the same or opposite eigenstate.  As the Hilbert space is thereby enlarged and contains non-physical states, ${K-N+D}$ constraints 
\begin{equation}\label{eq_constraint_hamiltonian}
    C_l=-\tilde \sigma_z^{(l_1)}\tilde \sigma_z^{(l_2)}\tilde \sigma_z^{(l_3)}[\tilde \sigma_z^{(l_4)}]
\end{equation}
on three or four qubits, represented by the last sum in $H_\text{parity}$ are introduced~\cite{lechner_quantum_2015, ender_parity_2023}. Here, $D$ denotes the number of ground state degeneracies of $H_\mathrm{op}$, and the square brackets in Eq.~\eqref{eq_constraint_hamiltonian} indicate that the fourth qubit involved in the interaction is optional. This mapping is illustrated in Fig.~\ref{fig_LHZ}(a).

The indices $l_i$ in Eq.~\eqref{eq_constraint_hamiltonian} are chosen such that all the logical indices involved in $C_l$ occur an even amount of times across all $l_i$. 
The constraints $C_l$ stabilize the space of logical states, i.e., of states that have a correspondence in $H_\mathrm{op}$ and are therefore valid. Crucially, the interactions $C_l$ can be implemented between qubits in geometrical vicinity on a two-diemansional (2D) grid with nearest-neighbor connectivity (see Fig.~\ref{fig_LHZ}(a)). Therefore, only geometrically local interactions are required for solving optimization problems of arbitrary order on digital or analog quantum devices, which is particularly important in the noisy intermediate-scale quantum (NISQ) era~\cite{Preskill2018quantumcomputingin} where implementing long-range interactions remains a challenge. Furthermore, the implementation of the constraint operators on digital quantum computers can be parallelized very efficiently. It was shown that the Parity Architecture is suitable for analog~\cite{lechner_quantum_2015} and digital quantum optimization~\cite{lechner_quantum_2020} as well as universal quantum computing~\cite{PhysRevLett.129.180503}. We note that the Parity Architecture can also be viewed as a error correction code for bit-flip errors, at the cost of implementing non-transversal logical $R_x$ rotations, resulting in non-local physical gates. However, for NISQ algorithms, such as the QAOA, we do not strive for full error correction, which is why we can exploit the Parity Architecture to \textit{remove} long-range interactions. Therefore, performing the Parity Transformation and adding the constraint operators to the resulting Hamiltonian allows for an implementation of Quantum annealing and the Quantum Approximate Optimization Algorithm~\cite{farhi2014quantum} with only local operations and, for the latter, in constant circuit depth.

\subsection{Parity QAOA\label{sec_parity_QAOA}}

The Quantum Approximate Optimization Algorithm (QAOA)~\cite{farhi2014quantum} aims at solving  optimization problems encoded in an $N$-qubit problem Hamiltonian $H_\mathrm{op}$ by preparing a solution candidate state
\begin{equation}
\ket{\psi(\bm{\beta}, \bm{\gamma})} = \prod_{j=1}^p U_x(\beta_j)U_\mathrm{op}(\gamma_j)\ket{+}
\end{equation}
for an optimization problem on $N$ qubits. Here, the initial state $|+\rangle$ denotes the state where all qubits $i$ are in an equal superposition of the eigenstates of $\sigma_z$, $(|0\rangle_i + |1\rangle_i)/\sqrt 2$, the unitary 
\begin{equation}
    U_\mathrm{op}(\gamma) = e^{-i\gamma H_\mathrm{op}}
\end{equation}
represents the time evolution operator under the problem Hamiltonian, and 
\begin{equation}
    U_x(\beta)=\prod_{i} e^{-i\beta \tilde \sigma_x^{(i)}}
\end{equation}
is the so-called driver unitary. The $2p$ parameters ${\bm{\beta}=(\beta_1, \dots, \beta_p)}$ and ${\bm{\gamma}=(\gamma_1, \dots, \gamma_p)}$ are optimized in a quantum-classical feedback loop by using a quantum computer to compute the energy expectation value ${\braket{H_\mathrm{op}}=\bra{\psi(\bm{\beta}, \bm{\gamma})}H_\mathrm{op}\ket{\psi(\bm{\beta}, \bm{\gamma})}}$, and employing a classical routine to optimize $\bm{\beta}$ and $\bm{\gamma}$ with respect to $\braket{H_\mathrm{op}}$ until some termination criterion is reached. The candidate ground state obtained as an optimization result is then determined by reading out all qubits.
By using the parity Hamiltonian $H_\text{parity}$ as the problem Hamiltonian and separating the local field term and the constraint term into two separate QAOA unitaries, extending the search space by ${\bm{\Omega}=(\Omega_1, \dots, \Omega_p)}$ to $3p$ classical parameters, parity QAOA~\cite{lechner_quantum_2020} is obtained.

Note that it is possible to implement the QAOA in the Parity Architecture without the constraint terms by allowing more complex and non-local driver operators~\cite{PRXQuantum.3.030304}. However, this approach exhibits a circuit depth growing linearly with the system size. In this work, we stick to the original proposal of explicitly implementing the $C_l$ terms in this work and we exploit the geometric locality of these operators. 

In its explicit form, parity QAOA requires single qubit operations on all qubits and an implementation of the three- and four-body constraints arising from the parity transformation, thus preparing the final state~\cite{lechner_quantum_2020}
\begin{equation}
    |\psi(\bm{\beta}, \bm{\gamma}, \bm{\Omega})\rangle = \prod_{j=1}^p U_z(\gamma_j) U_c(\Omega_j) U_x (\beta_j) |+\rangle,
\end{equation}
where the unitary
\begin{equation}
    U_z (\gamma) = e^{ -i \gamma \sum_i \tilde J_i \tilde \sigma_z^{(i)} }
\end{equation}
is the time evolution under the problem Hamiltonian after the parity transformation. The operator
\begin{equation}
    U_c (\Omega) = e^{ -i \Omega \sum_l \tilde \sigma_z^{(l_1)} \tilde \sigma_z^{(l_2)} \tilde \sigma_z^{(l_3)} \tilde \sigma_z^{(l_4)} }
\end{equation}
enforces the constraints on all plaquettes $l$. This total time evolution is repeated for a number of rounds $p$.

\begin{figure}
    \centering
    \includegraphics[width=0.49\textwidth]{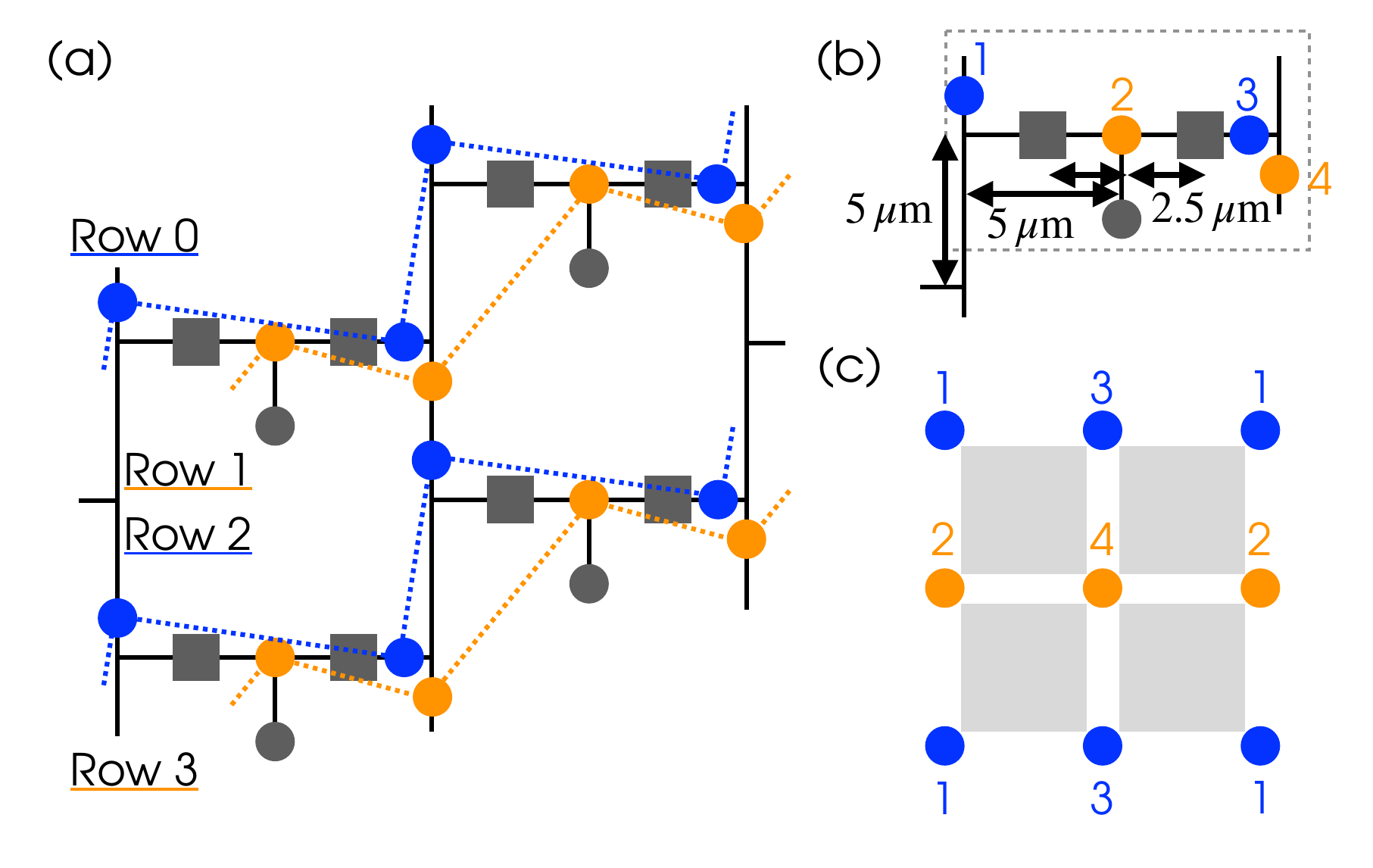}
    \caption{The spin bus platform under consideration~\cite{ArqueProposal}. (a) The electrons (blue and orange dots) are stored in shuttling lanes (black lines), qubits are labeled 1-4 in each unit cell. The electrons can be conveyed to manipulation (dark gray squares) or initialization/readout zones (dark gray circles), T-junctions provide connectivity between qubits in two dimensions. The vertical shuttling lanes are controlled globally. The segment holding qubits 1-4 can be considered a unit cell of the lattice. For the purpose of the QAOA, the electrons are separated in even (blue) and odd (orange) rows indicated by the dashed lines. They correspond to the rows from Fig.~\ref{fig_LHZ}.
    (b) One unit cell with four qubits, highlighted by the dashed box. We assume that qubits 1, 3 and 4 are placed $\SI{1.25}{\mu\meter}$ from the T-junction (and the manipulation zone, in case of qubit 3). The classical control electronics and the gate fan-out can find space between the unit cells. 
    (c) Corresponding plaquettes of four-qubit constraints (light gray) of the Parity Architecture. The labels denote the qubits' position in their unit cell.}
    \label{fig_spinbus}
\end{figure}

The operators $U_x$ and $U_z$ can be trivially implemented with single qubit rotations, and it has been shown~\cite{lechner_quantum_2020, unger_low-depth_2022} that also $U_c$ can be implemented in constant depth. We choose the decomposition from Ref.~\cite{unger_low-depth_2022}, which is shown in Fig.~\ref{fig_LHZ}(c), to achieve the latter. This approach separates the square grid of qubits into ribbons of two neighboring rows each. The constraints are then enforced by a sequence of CNOT and ZZ gates which can be executed in parallel on each ribbon and furthermore allows parallelization of all ribbons that include an even-numbered top row of qubits in the first step and ribbons with odd top rows in the second steps. The rows are indicated in Figs.~\ref{fig_LHZ}, \ref{fig_spinbus} and \ref{fig_modular}. In this work, for simplicity, we consider the case with an implementation of all constraints in the sequence of eight time steps from Fig.~\ref{fig_LHZ}(c), where at each boundary between two plaquettes both qubits are included in the respective constraints~\cite{lechner_quantum_2015} (i.e., square plaquettes only). More general problems~\cite{ender_parity_2023} can be treated by including additional CNOT gates between steps 2 and 3~\cite{unger_low-depth_2022}.

Apart from the advantages of solely local interactions and a constant circuit depth for QAOA, the Parity Architecture offers the advantage of an intrinsic possibility for error mitigation. The redundant information held by the additional qubits can be exploited to read out several spanning trees of the logical system in order to detect and correct constraint violations, either due to quantum errors or because of constraint violations by the mixer unitary~\cite{PhysRevA.108.032408}. The latter can also be avoided by using constraint-preserving driver operators~\cite{PRXQuantum.3.030304}, however at the cost of giving up the constant circuit depth. Even though physical errors reduce the success rate for finding the ground state, it was shown that the decoding of the spanning trees still grants a high success rate and can outperform standard QAOA~\cite{PhysRevA.108.032408}.

\section{Parity QAOA with Spin Qubits\label{sec_architectures}}

In this section we present an implementation of the Parity QAOA on two possible two-dimensional electron spin qubit architectures: an extremely sparse architecture entirely based on spin shuttling~\cite{ArqueProposal} in Sec.~\ref{sec_spin_bus} and an architecture with small, dense registers of QDs connected by shuttling links~\cite{Vandersypen2017} in Sec.~\ref{sec_modular}. Here, the mapping to the chip layout and the compiled shuttling and gate sequences are given, a comparison of the performances will follow in Sec.~\ref{sec_error_evaluation}. Animated versions of the gate and shuttling sequence can be found in the Supplemental Material~\cite{supp}.

\subsection{Implementation on a Sparse Spin Bus Architecture\label{sec_spin_bus}}

Recently, the so-called spin bus architecture, Fig.~\ref{fig_spinbus}, was proposed~\cite{ArqueProposal}. There, the electrons are stored in shuttling lanes defined by periodically interconnected voltage gates, which allow a smooth conveyor-mode shuttling of the charge by applying phase shifted sinusoidal voltages to each set of connected gates~\cite{Seidler2022,Struck2023,Xue2023}. The shuttling lanes form a two-dimensional lattice with dedicated manipulation and initialization/readout zones to which the electrons can be shuttled on demand. This architecture has the advantage of high connectivity and promises a long coherence time, since the electrons can be stored far from the detrimental effects of the micromagnets at the manipulation zones. Due to its sparse nature, it can be expected to suffer from little crosstalk and it creates space for the voltage gates and classical control electronics. The architecture is sketched in Fig.~\ref{fig_spinbus}.

Unlike Ref.~\cite{ArqueProposal}, we assume that each unit cell contains four electrons and two manipulation zones, doubling the number of qubits at the cost of the additional input lines for one conveyor per cell. Thus, the number of qubits is doubled with only a moderate increase in complexity. We expect this to be beneficial for near-term devices limited by the requirements of room-temperature control. Furthermore, the increased density of qubits results in shorter shuttling paths within the unit cells, thus mitigating shuttling-related errors. Alternatively, it would also be possible to  carry out the operations sequentially while storing electrons not involved in shuttling lanes, thus reducing complexity at the cost of a slightly reduced performance.

\begin{figure}
    \centering
    \includegraphics[width=0.49\textwidth]{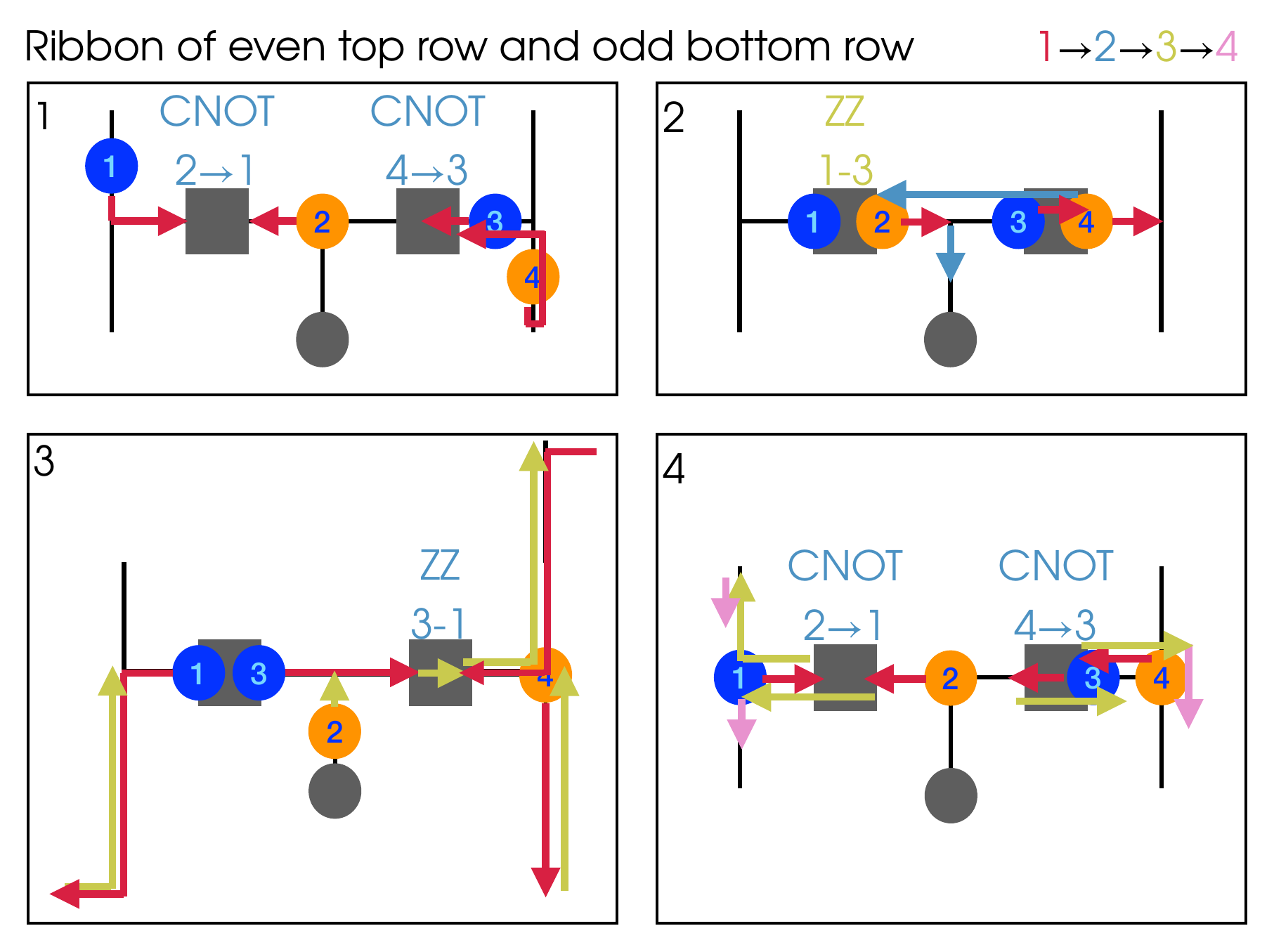}
    \caption{Circuit for implementing the parity constraints on ribbons of an even top and an odd bottom row if executed on each unit cell in parallel. The arrows indicate the operation of spin conveyors, gate operations are indicated near the manipulation zone. The coloring represents the time ordering of the sequence: from red to light blue to lime to pink as indicated in the top right corner. Due to the global vertical shuttling lanes, some electrons are moved unnecessarily, for example qubit 4 in steps 1 and 3 and qubit 1 in step 4. Note that qubit 1 leaves its unit cell and undergoes a gate in a different cell in step 3, then returns in step 4, such that the second ZZ gate is between qubits from different unit cells. Steps 1-4 correspond to steps 1-4 in Fig.~\ref{fig_LHZ}(c).}
    \label{fig_spinbs_even-odd}
\end{figure}

\begin{figure}
    \centering
    \includegraphics[width=0.49\textwidth]{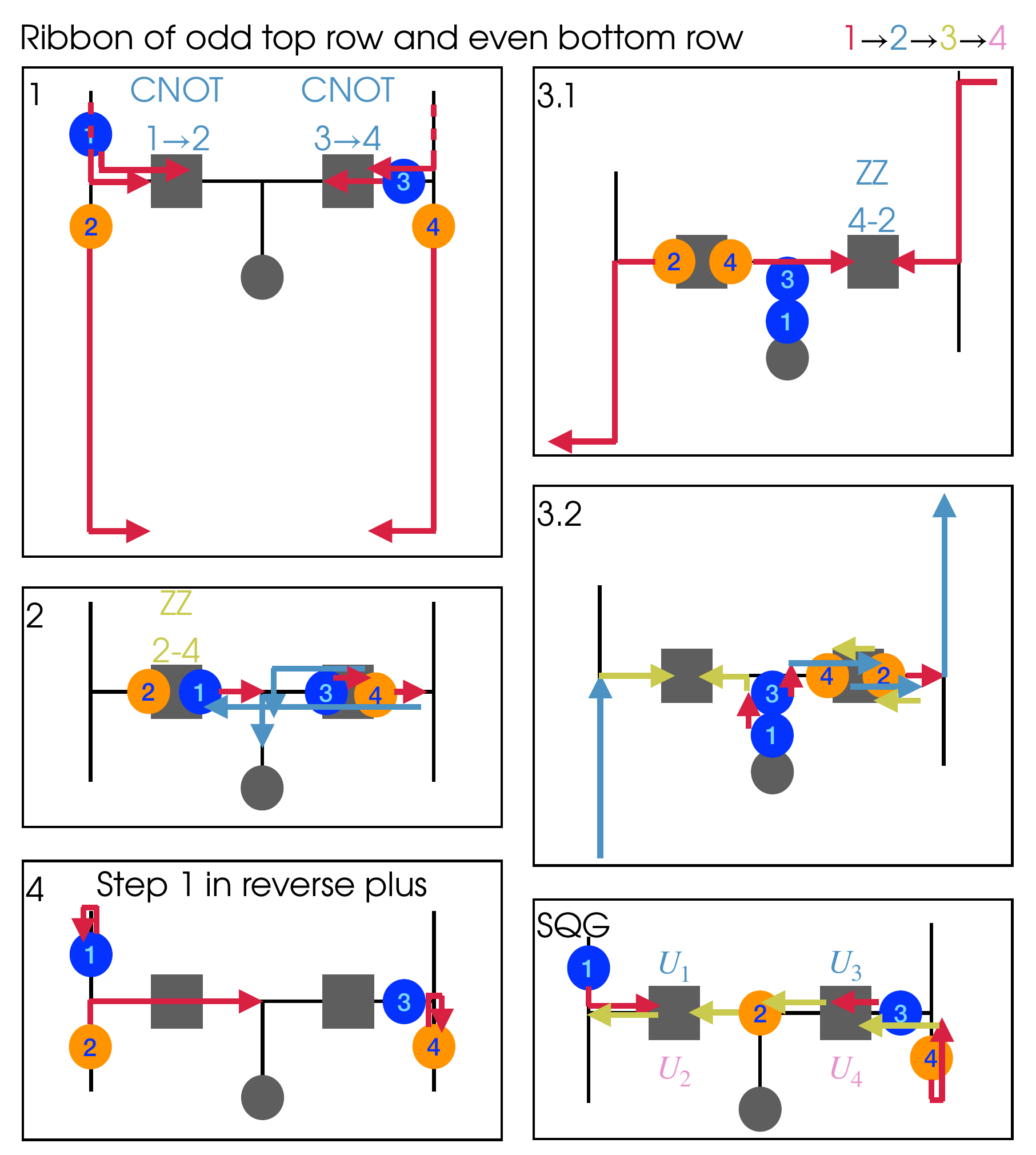}
    \caption{Circuit for implementing the parity constraints on ribbons of an odd top and an even bottom row (1-4) if executed on each unit cell in parallel, as well as for single-qubit gates (SQG). In order to implement the constraints on this set of ribbons qubits 2 and 4 are shuttled to an adjacent unit cell in step 1 and return in step 4. The third step is separated in 3.1, which includes the gate, and step 3.2, which restores the final configuration of step 1 such that the final CNOT gate and the shuttling in step 4 can be performed easily.
    The circuit SQG for implementing arbitrary single-qubit gates $U_i$ on all qubits $i$ can be followed by returning the qubits to their home position (initial configuration of step 1 in Fig.~\ref{fig_spinbs_even-odd}), or by returning qubits 1 and 3 to the manipulation zone (final configuration of step 1 in even-odd ribbons) if the single qubit gate is followed by a constraint step.}
    \label{fig_spinbs_odd-even}
\end{figure}

The constraints are implemented by the gate sequence shown in Fig.~\ref{fig_spinbs_even-odd} for the set of ribbons consisting of even top and odd bottom rows, followed by the circuit in Fig.~\ref{fig_spinbs_odd-even} for the set of ribbons consisting of odd top and even bottom rows. Due to the non-equivalent positions of the qubits in the unit cell, these two cases need to be treated separately. The shuttling and gate operations can be performed in parallel on all unit cells, achieving an optimal circuit depth. Due to the global control of the vertical shuttling lanes, some electrons are shuttled unnecessarily, as shown in Fig.~\ref{fig_spinbs_even-odd}. However, with high-fidelity shuttling sufficient for several tens of micrometers, we do not expect this short excess distance to be problematic, averaging quasistatic noise by shuttling (motional narrowing) might even enhance the coherence of an idle qubit.

During the constraint circuit, every bulk qubit -- which is not located on the edge of the physical device -- is target and control of two CNOT gates each and participates in two ZZ gates while qubits at the boundary of the lattice experience fewer gates. Qubit 1 is shuttled over a distance of $\SI{42.5}{\mu\meter}$ with our assumptions on the size of the unit cell and idles for $\SI{46.25}{\mu\meter}/v + 2T_{ZZ}$ where $v$ is the shuttling velocity and $T_{ZZ}$ is the time required for a ZZ gate. Qubit 2 is shuttled over $\SI{62.5}{\mu\meter}$ and idles for $\SI{26.25}{\mu\meter}/v + 2T_{ZZ}$, qubit 3 is shuttled over $\SI{26.25}{\mu\meter}$ and idles for $\SI{62.5}{\mu\meter}/v + 2T_{ZZ}$, while qubit 4 moves by $\SI{61.25}{\mu\meter}$ and idles for $\SI{27.5}{\mu\meter}/v + 2T_{ZZ}$. This is not counting the shuttling of step 1 for the even-odd ribbons since it can be absorbed into the single-qubit gates of the driver term. The shuttling velocity $v$ is a critical parameter for the duration of the algorithm and also for the strength of the errors, as will be discussed in Sec.~\ref{sec_error_model}. The discrepancy in shuttled distance between the qubits is due to the fact that qubits 2 and 4 are moved to a foreign unit cell in order to implement the constraints on the odd-even ribbons, while qubit 3 is favorably placed close to two manipulation zones and is thus highly connected without much movement.

The circuit for the execution of single-qubit gates, required for implementing $U_x$ and $U_z$, is shown in Fig.~\ref{fig_spinbs_odd-even} in the panel SQG. In a single qubit gate step followed by the constraint circuit, qubits 1 and 3 can be returned to the manipulation zone after the operation on qubits 2 and 4 is complete, thus simplifying step 1 from the even-odd rows, Fig.~\ref{fig_spinbs_even-odd}. Alternatively, all qubits can be returned to their idling position, e.g., to wait until they are read out at the end of the algorithm. The former version adds $\SI{8.75}{\mu\meter}$ ($\SI{2.5}{\mu\meter}$, $\SI{6.25}{\mu\meter}$, $\SI{6.25}{\mu\meter}$) of shuttling on qubit 1 (2, 3, 4) while the latter adds $\SI{10}{\mu\meter}$ ($\SI{5}{\mu\meter}$, $\SI{7.5}{\mu\meter}$, $\SI{10}{\mu\meter}$) on qubit 1 (2, 3, 4).

Initialization and readout, the remaining building blocks of any algorithm, are performed by successively shuttling each qubit from or to the initialization/readout zone. In order to initialize each qubit to an arbitrary state, electrons 1, 3 and 4 can be stopped at the manipulation zone they pass on the way to their starting position, and qubit 2 can be routed on a detour to either manipulation zone.

\subsection{Implementation on a Modular Architecture\label{sec_modular}}

\begin{figure}
    \centering
    \includegraphics[width=0.49\textwidth]{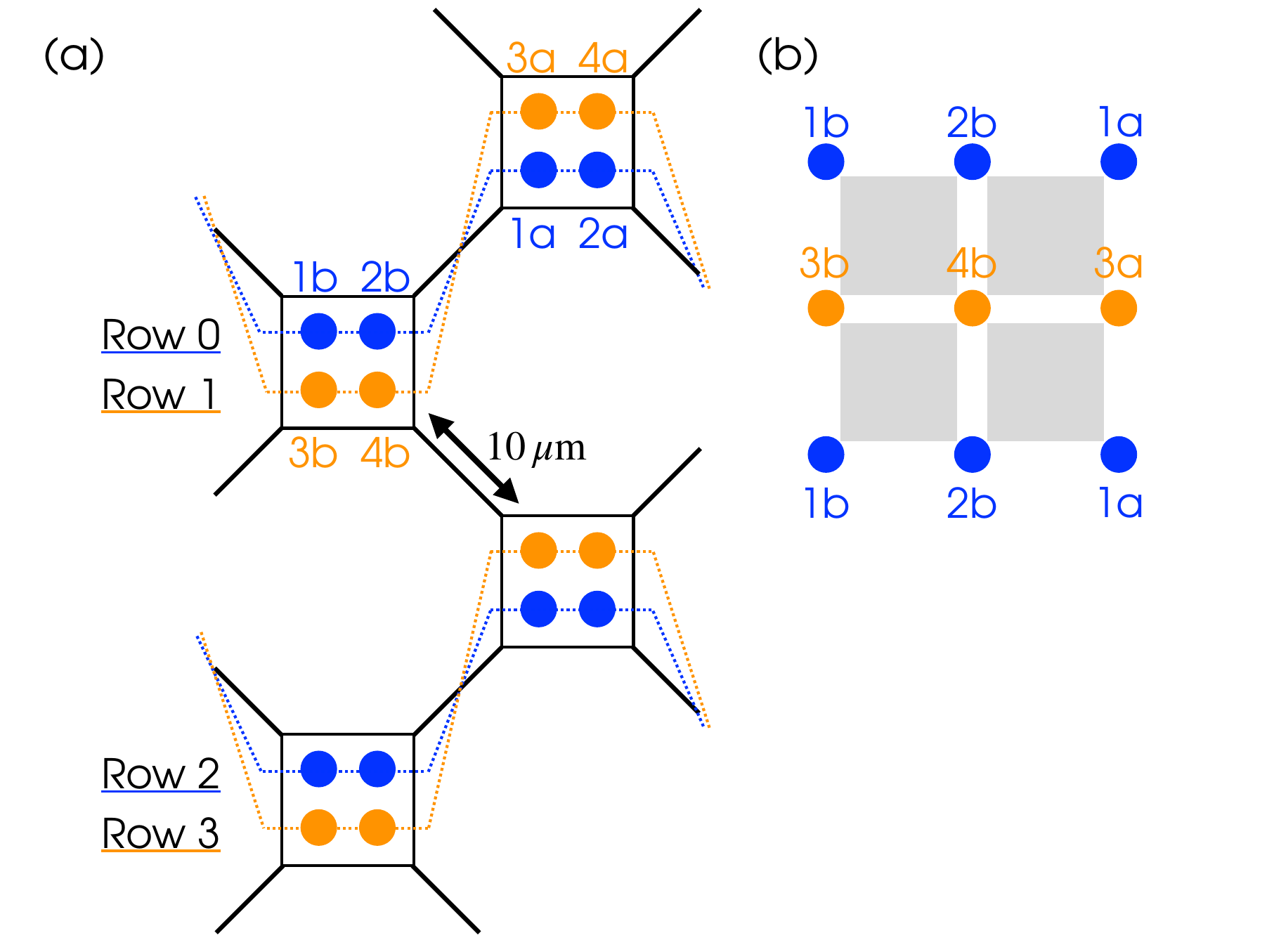}
    \caption{The modular architecture under consideration~\cite{Vandersypen2017}. (a) Registers of ${2\times 2}$ QDs are interconnected via diagonal shuttling lanes. Manipulations are performed on-site and we assume that one site per register is equipped for readout, shuttling serves for connectivity only. Here, a unit cell consists of two registers denoted by a and b with eight qubits in total, since it is favorable to mirror the registers alternately. This architecture provides space for classical electronics and the gate fan-out in between the registers.
    (b) Corresponding plaquettes of the Parity Architecture. The labels denote the qubits' position in their unit cell.}
    \label{fig_modular}
\end{figure}

A trade-off between the typical dense array of QDs~\cite{Borsoi2023} and the extremely sparse spin bus is an architecture where dense registers of a few qubits are connected by coherent quantum links, as depicted in Fig.~\ref{fig_modular}~\cite{Vandersypen2017,Crawford2023}. This retains the advantages of dense arrays -- fast gates and a small footprint -- while creating space for the fan-out of voltage gates and classical electronics. Here, we consider a minimal version of this modular architecture, where registers of $2\times 2$ QDs are connected via diagonal conveyor-mode shuttling lanes in two dimensions as depicted in Fig.~\ref{fig_modular}.

\begin{figure}
    \centering
    \includegraphics[width=0.49\textwidth]{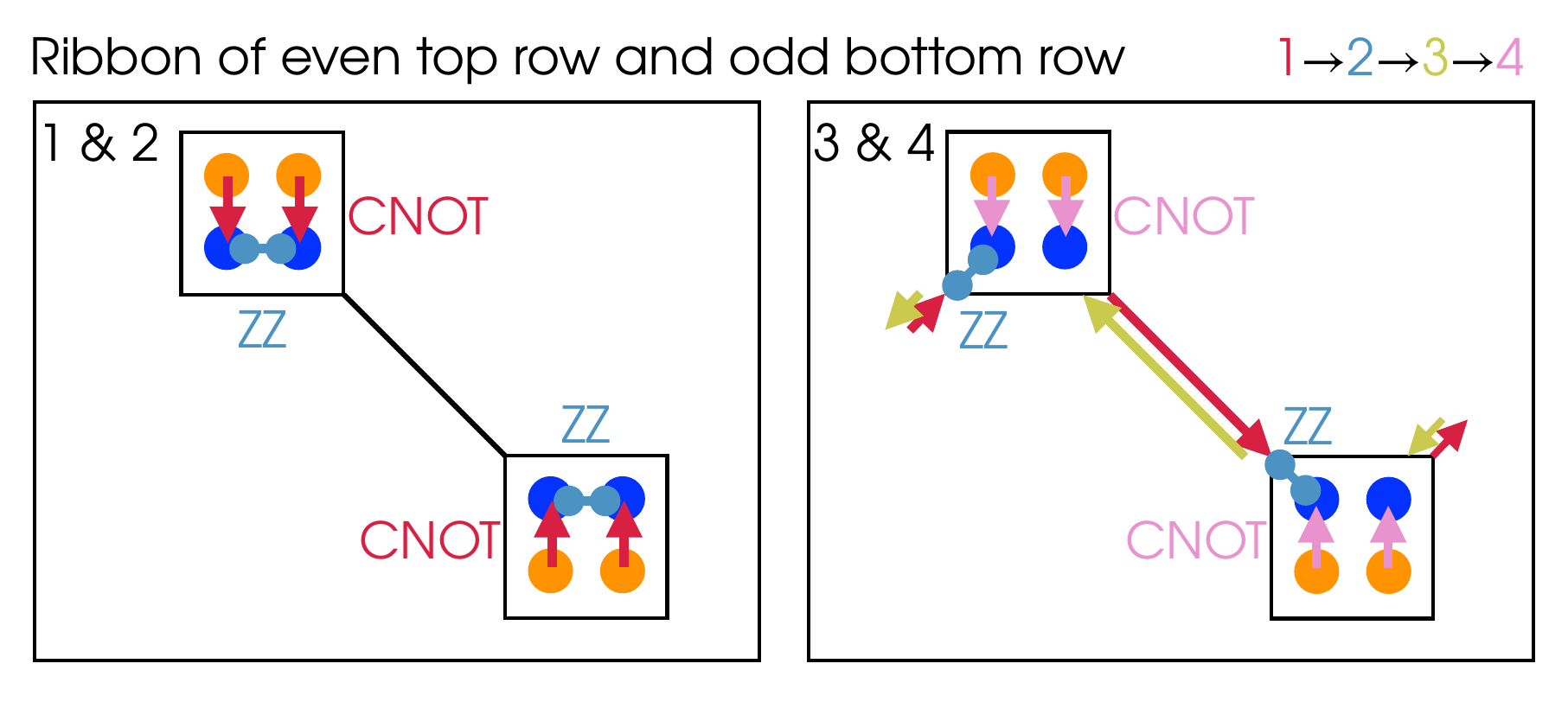}
    \caption{Circuit for implementing the parity constraints on ribbons of even top and odd bottom rows in a modular architecture if executed on each unit cell in parallel. Arrows and lines within the registers indicate two-qubit gates, arrows between the registers indicate the operation of shuttling lanes. The time ordering is indicated by the colors, from red to light blue to lime to pink as indicated in the top right corner. The steps 1-4 here correspond to the decomposition of the circuit in Fig.~\ref{fig_LHZ}(b) and on the spin bus in Fig.~\ref{fig_spinbs_even-odd}.}
    \label{fig_modular_even-odd}
\end{figure}

\begin{figure}
    \centering
    \includegraphics[width=0.49\textwidth]{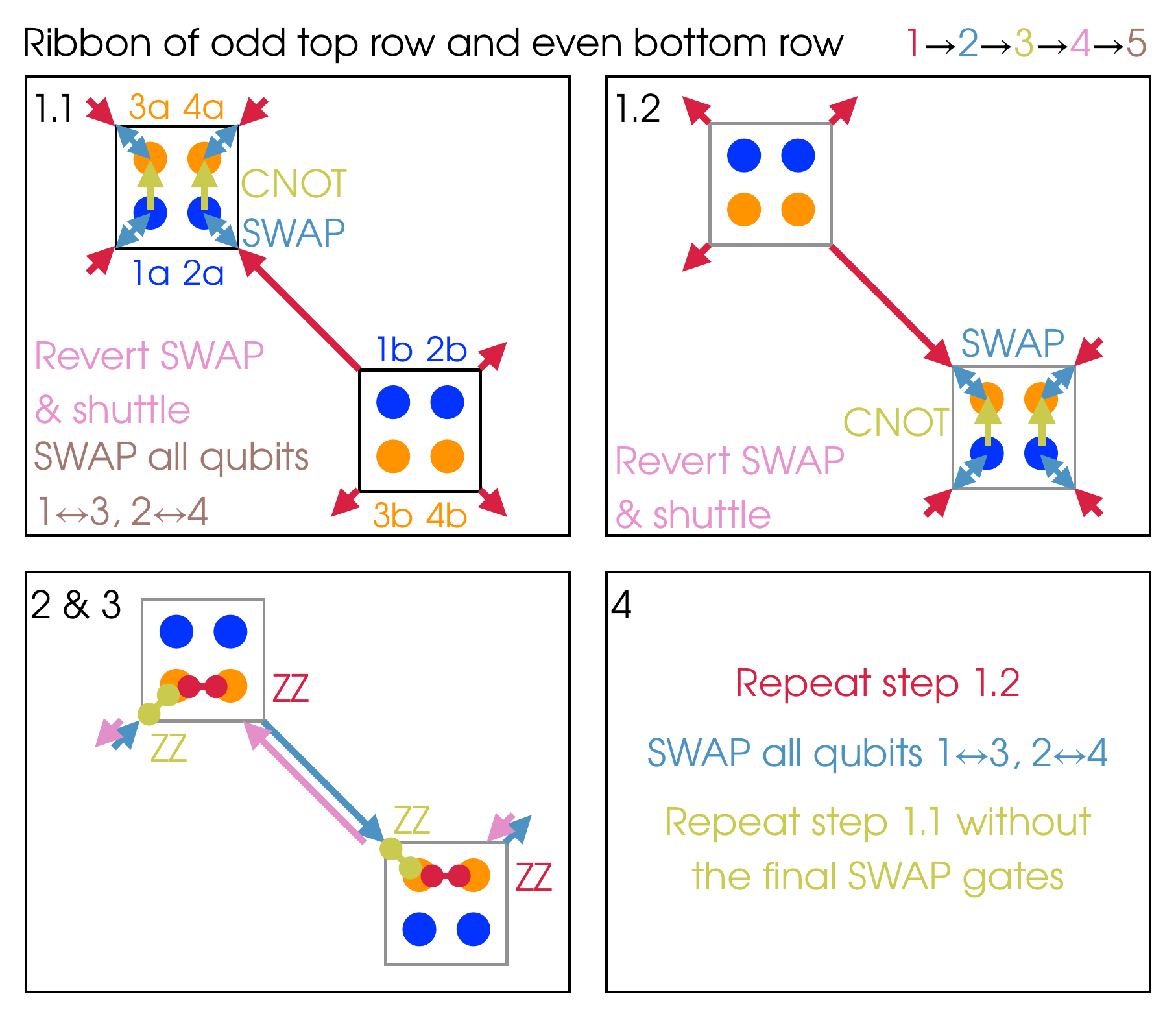}
    \caption{Circuit for implementing the parity constraints on ribbons of an odd top and an even bottom row in a modular architecture if executed on each unit cell in parallel. Here, the lower connectivity compared to the spin bus is apparent, as a each qubit is subject to a total of 10 SWAP gates in order to implement the required interaction. The steps 1-4 correspond to those in Fig.~\ref{fig_LHZ}(b) and the implementation on the spin bus in Fig.~\ref{fig_spinbs_odd-even}. In steps 1 and 4, the pulses for the gates can be applied globally to all registers since all QDs in the non-addressed register are vacant. Note that in steps 1 and 4 the qubits 1a-4a and 1b-4b in the mirrored registers undergo the same operations, albeit in different order.}
    \label{fig_modular_odd-even}
\end{figure}

A sequence of operations for implementing the constraint term on this architecture is shown in Figs.~\ref{fig_modular_even-odd} and \ref{fig_modular_odd-even}. Here, a number of SWAP operations cannot be avoided since each qubit is coupled to only three direct neighbors, which is not ideal for realizing the Parity Architecture. Again, we find that for the constraint term all bulk qubits are control and target of two CNOT gates each, and they participate in two ZZ gates and, additionally, ten SWAP gates. Furthermore, all qubits at positions 1 and 3 in the unit cells are shuttled by $\SI{40}{\mu\meter}$ and idle for $\SI{80}{\mu\meter}/v + 2 T_\mathrm{ZZ} + 2 T_\mathrm{CNOT}$, while all qubits at positions 2 and 4 are shuttled by $\SI{60}{\mu\meter}$ and idle for $\SI{60}{\mu\meter}/v + 2 T_\mathrm{ZZ} + 2 T_\mathrm{CNOT}$.  It is possible to replace eight SWAP gates per qubit with hopping between neighboring QDs within the registers, if an alternative circuit is used and empty registers are available at the edges of the lattice (see Appendix~\ref{app_AltMod}).

Single-qubit gates can be performed on-site in each QD and do not require shuttling, and we assume that one QD per register is equipped with a readout apparatus, allowing us to measure and initialize all qubits in four successive steps combined with SWAP gates or other operations to transfer the spin projection between neighboring QDs. In summary, time and operations are comparable on both architectures, with the addition of 10 SWAP gates in the modular architecture. The spin bus is able to implement a single round of QAOA $\approx\SI{10}{\mu\meter}/v$ faster than the modular architecture, including initialization and readout.

\section{Gate and Error Model\label{sec_error_model}}

In this section we introduce our error model, which is applied to both hardware layouts. There are three general sources of errors: 
\begin{enumerate}[label=(\roman*)]
    \item Errors from the quantum operations on the qubits. 
    \item Errors from the shuttling processes.
    \item Errors from initialization and readout.
\end{enumerate}
For the specific details of the hardware platform, we assume electron spins confined in QDs in a Si/SiGe heterostructure. This choice is motivated by the outstanding coherence properties of QDs in isotopically purified silicon~\cite{RevModPhys.95.025003}, the demonstration of high-fidelity quantum gates~\cite{Xue2022,Mills2022,Noiri2022} and the recent successes of electron conveyors in this material~\cite{Struck2023,Xue2023}. Thus, all necessary building blocks for our architecture are available.

\subsection{Gate Errors}

We assume that each manipulation zone in the spin bus architecture and each register in the modular architecture is equipped with a micromagnet. Thus, single spin manipulation can be accomplished by means of electric dipole spin resonance. Lowering the tunnel barrier between two adjacent QDs gives rise to a nearest neighbor exchange interaction~\cite{RevModPhys.95.025003}. We decompose all two-qubit gates into gates that are well-proven and optimized~\cite{RevModPhys.95.025003,Mills2022,Xue2022}.

The gate errors are modeled using the Kraus representation of imperfect quantum channels~\cite{NielsenChuang} where the error probabilities are chosen in order to describe realistic error rates. For single-qubit gates, we assume a depolarization channel with a probability of $p_d = 10^{-3}$~\cite{Xue2022,Mills2022,Noiri2022,ArqueProposal}, thus the density operator $\rho_f$ after a single qubit gate $U_i$ on qubit $i$ is given by $\rho_ f = \sum_k K_{k,i} \rho_0 K_{k,i}^\dagger$, where ${K_{0,i} = \sqrt{1-p_d}\, U_i}$ and ${K_{k,i} = \sqrt{p_d/3}\, \tilde \sigma_k^{(i)}}$, for ${k = x,y,z}$ with the Pauli operators $\tilde\sigma_k^{(i)}$ of qubit $i$ and the input density matrix $\rho_0$.

The default entangling two-qubit gate in QDs equipped with a micromagnet is the controlled phase gate, ${\mathrm{CP}_\alpha = \mathrm{diag}(1,1,1,e^{i\alpha})}$, in the basis $\{ |0,0\rangle, |0,1\rangle, |1,0\rangle, |1,1\rangle \}$~\cite{Xue2022,Mills2022}. Similar to the single-qubit gates, we assume that the errors of the CP gate are captured by a phase flip channel with probability $p_\phi = 10^{-3}$ and a bit flip channel with probability $p_b = 10^{-6}$ on both qubits, taking into account that dephasing is the limiting error mechanism for spin qubits. However, the circuits presented in the previous section require the two-qubit gates ZZ, CNOT and SWAP. These are synthesized from CP and single-qubit gates by concatenating the respective quantum channels.

The gate ZZ$^{(i,j)}_\alpha = \exp(i \alpha \tilde\sigma_z^{(i)} \tilde\sigma_z^{(j)})$ between qubits $i$ and $j$ with rotation angle $\alpha$ is obtained from the decomposition 
\begin{equation}
\mathrm{ZZ}^{(i,j)}_\alpha = \mathrm{CP}_{-2\alpha}^{(i, j)} R_z^{(i)}(\alpha) R_z^{(j)}(\alpha)
\end{equation}
with the single qubit rotation $R_z^{(i)}(\alpha) = e^{-i \alpha \tilde\sigma_z^{(i)}/2}$ around the $z$ axis of qubit $i$. Note that the ZZ gate and its error are symmetric between the two qubits.

Analogously, the CNOT gate with control qubit $i$ and target $j$ can be represented as 
\begin{equation}
    \mathrm{CNOT}^{(i,j)} = H^{(j)} \mathrm{CP}^{(i,j)}_\pi H^{(j)}
\end{equation} 
with the Hadamard gate $H$. A SWAP gate is then obtained by a sequence of three CNOTs: 
\begin{equation}
    \mathrm{SWAP}^{(i,j)} = \mathrm{CNOT}^{(j,i)} \mathrm{CNOT}^{(i,j)} \mathrm{CNOT}^{(j,i)}.
\end{equation}
These composite gates can reliably be performed with a high fidelity in Si/SiGe quantum dots with micromagnets~\cite{Xue2022,Mills2022}, although we expect that it is possible to engineer a more efficient version of the SWAP gate realized by the exchange coupling or by including a physical position swap via shuttling.

Qubits idling for a time $t$ will suffer from dephasing with a characteristic timescale $T_2$ and relaxation with a characteristic timescale $T_1$ due to environmental noise. A conservative lower bound for $T_2$ is the pure dephasing time $T_2^*$. These are modeled with the Kraus representation of a phase phase damping and amplitude damping channel~\cite{NielsenChuang}. For ${t\ll T_2,T_1}$ the probabilities for these channels can be approximated as ${p_{\phi,\mathrm{idl}} = t/T_2}$ (${p_{r,\mathrm{idl}} = t/T_1}$) for dephasing (relaxation). If the dephasing is dominated by quasistatic noise, the decay of the coherences is described by a Gaussian by the choice ${p_{\phi,\mathrm{idl}} = (t/T_2)^2}$~\cite{RevModPhys.95.025003}.

As optimistic estimate we use $T_1 = \SI{1}{\second}$ for both architectures and ${T_2 = \SI{20}{\micro\second}}$ for the modular layout~\cite{Struck2020}, although this number can improve significantly if dynamical decoupling is applied~\cite{RevModPhys.95.025003}. The spin bus allows storing the electrons further away from the detrimental field of the micromagnet, which couples the spin to electric field fluctuations, as well as the SETs and charge reservoirs, which are sources of Johnson noise. Thus a longer dephasing time can be expected. Using SiMOS QDs without micromagnet as a reference we take ${T_2 = \SI{100}{\micro\second}}$ as an optimistic estimate for the spin bus layout~\cite{RevModPhys.95.025003}. Note that in both architectures, shuttling has an effect similar to dynamical decoupling and can increase a qubit's coherence time due to motional narrowing~\cite{PRXQuantum.4.020305,Struck2023}. This is particularly relevant if a qubit is shuttled back and forth to and from a manipulation zone along the same path: Inverting the qubit state before the return allows the removal of certain adiabatic effects, such as deterministic variations of the qubit frequency during the shuttling~\cite{PRXQuantum.4.020305}. This is trivially done in the modular architecture and is possible for most shuttling paths in the spin bus. Idling mostly occurs while qubits are waiting for the completion of shuttling processes or gates on other qubits. Typical timescales are ${T_{1q} = \SI{100}{\nano\second}}$ for the duration of a single qubit gate and ${T_{2q} = \SI{50}{\nano\second}}$ for the duration of a native two-qubit gate~\cite{ArqueProposal}.

\subsection{Shuttling Errors}

In both architectures we assume conveyor-mode shuttling of electrons and we have to take into account that the chosen host material exhibits near-degenerate valleys~\cite{RevModPhys.54.437,PhysRevLett.88.027903,PhysRevB.81.115324}. 
A detailed derivation of shuttling errors can be found in Ref.~\cite{PRXQuantum.4.020305}, however, we extend the model for the effects of the valley pseudospin in order to take into account the recent discoveries concerning the dependence of the valley splitting on the alloy disorder~\cite{PaqueletWuetz2022,Lima_2023_1,Lima_2023_2}.

Both, valley splitting $E_v$ and valley phase $\varphi$ will follow a random trajectory along the shuttling path. Assuming that spin-valley relaxation hot spots~\cite{PhysRevB.72.155410,PhysRevB.71.205324,PhysRevLett.110.196803} are avoided, the main effect of the valley is dephasing~\cite{PRXQuantum.4.020305}. Due to non-adiabatic shuttling the electron may have a random weight in the excited valley state and thus accumulate an unpredictable phase. To capture this effect, we assign each location $x$ along the shuttling path a valley Hamiltonian
\begin{equation}
    H'_v (x) = \frac{E_v(x)}{2} \left( e^{i \varphi (x)} |v_-\rangle\langle v_+| + e^{-i \varphi (x)} |v_+\rangle\langle v_-| \right),
\end{equation}
with the $\pm z$ valley states $|v_\pm\rangle$~\cite{PhysRevB.81.115324}. At each point $x$ the valley splitting and phase are drawn from distributions $P_{E_v}$ and $P_\varphi$ respectively.

Since the position is time-dependent, $x = vt$ with shuttling velocity $v$, transforming $H'_v$ into its instantaneous eigenbasis with the unitary $U$ leads to a term that causes transitions between the instantaneous valley eigenstates,
\begin{eqnarray}
    H_v &=& U^\dagger H'_v U + i \hbar \dot{U}^\dagger U\\
    &=& \frac{E_v}{2} \tau_z + \frac{\hbar \dot \varphi}{2} (1\!\! 1 - \tau_x),
\end{eqnarray}
where $\tau_{z} = |e_v\rangle\langle e_v| - |g_v\rangle\langle g_v|$ and $\tau_{x} = |g_v\rangle\langle e_v| + |e_v\rangle\langle g_v|$ are the Pauli $z$ and $x$ matrices for the instantaneous valley eigenstates. Approximating the differentiation as a quotient results in $\dot \varphi (t) = \Delta \varphi / \Delta t = v \Delta\varphi / \Delta x$, where $\Delta x$ is the minimal distance at which a different valley splitting is resolved, and $\Delta \varphi$ is the difference in valley phase over this distance. A reasonable assumption for $\Delta x$ is the dot size~\cite{vollmer2024}. The difference $\Delta \varphi$ is a random variable whose probability distribution function can be obtained from a convolution of the distribution functions of the summands,
\begin{equation}
    P_{\Delta\varphi} = \int_{-\pi}^\pi \mathrm{d}\phi\, P_\varphi(\phi) P_\varphi [- (\Delta\varphi - \phi)].
\end{equation}

In the limit of a large valley splitting relative to the variation of the valley phase, ${E_v/\hbar\dot{\varphi}}\gg 1$, the probability of finding the electron in its excited valley state after moving it over a distance of $\Delta x$ is easily obtained from the solution of the Schr\"odinger equation for a two-level system in the adiabatic frame~\cite{PRXQuantum.4.020305},
\begin{eqnarray}
    p_{e,v} &=& |\langle e_v | e^{-i H_v \Delta x / \hbar v} | g_v \rangle|^2\\
    &=& \frac{(\hbar v \Delta\varphi / \Delta x)^2}{E_v^2 + (\hbar v \Delta\varphi / \Delta x)^2} \sin^2\left(\theta\right), \\
    \theta &=&  \sqrt{E_v^2 + (\hbar v \Delta\varphi / \Delta x)^2} \frac{\Delta x}{2\hbar v}.
\end{eqnarray}
Given the probability distributions, the average excitation probability over the distance $\Delta x$ is thus
\begin{equation}
    \bar{p}_{e,v} = \int_0^\infty \mathrm{d} E_v \int_{-\pi}^\pi \mathrm{d} (\Delta \varphi)\, P_{E_v}(E_v) P_{\Delta \varphi}(\Delta\varphi) p_{e,v}.\label{eq_mean_valley_excitation}
\end{equation}
Consequently, the average probability for finding the electron in the excited valley state after moving a distance ${L = n \Delta x}$ is given by 
\begin{equation}
p_v = 1 - (1 - \bar{p}_{e,v})^n \approx \bar{p}_{e,v} L / \Delta x.    
\end{equation}

In Ref.~\cite{Lima_2023_2}, a Rice distribution was found for the valley splitting $E_v$ and the numerical results of the same reference suggest ${P_{\varphi} \approx 1/2\pi}$ and thus ${P_{\Delta\varphi} \approx 1/2\pi}$. Consequently, $p_v$ only depends on the two parameters of the Rice distribution -- corresponding to the mean valley splitting and its variance -- the shuttling velocity, dot size and traveled distance. Our model for $p_v$ is easily adapted to more accurate distribution functions unveiled by future research and hardware-specific distributions measured from individual devices by simply evaluating Eq.~(\ref{eq_mean_valley_excitation}).

In accordance with Ref.~\cite{PRXQuantum.4.020305} we then include an adiabatic contribution 
\begin{equation}
p_\mathrm{ad} = 2 l_c^{\delta\omega} L / (v T_2)^2
\end{equation}
to the dephasing, due to fluctuations of the spin splitting with the correlation length $l_c^{\delta\omega}$. The fluctuations of the spin splitting can originate from surrounding nuclear spins, magnetic field gradients, and spin-orbit coupling. This expression accounts for motional narrowing, which partly protects the shuttled spin. However, this effect is expected to be reduced in isotopically purified silicon, since the electric fluctuations dominating the noise in the absence of nuclear spins have a comparably large correlation length. Note that the effect of deterministic and reproducible variations of the spin splitting can be removed by calibration.

We choose a dot size of ${\Delta x = \SI{20}{\nano\meter}}$ and estimate a noise correlation length of ${l_c^{\delta\omega} = \SI{1}{\mu\meter}}$. The latter is very conservative, but the effect of motional narrowing is strongly suppressed in that order of magnitude already. The dephasing during the shuttling is then described by the Kraus representation of a dephasing channel with the probability 
\begin{equation}
p_\text{deph} = 1 - (1 - p_v) (1 - p_\mathrm{ad}) \approx p_v + p_\mathrm{ad}.    
\end{equation}
 
Relaxation of a shuttled spin is described by an amplitude damping channel with the probability ${L / v T_1 + 10^{-4} L /\SI{10}{\micro\meter}}$, where the second term emerges due to spatially varying transverse spin orbit components~\cite{PRXQuantum.4.020305}.

Note that after averaging the valley effects this is only the expected shuttling error in a typical shuttling lane. Due to its probabilistic nature, the valley splitting may strongly fluctuate between different shuttling lanes on a given device. Thus, all fidelity estimates obtained from this model represent an expectation value averaged over a large number of devices.

A number of possible shuttling errors are neglected in this description. These include the temporary breaking of a moving quantum well into a double well which may harm the orbital state of the electron, the loss or capture of an electron, and the jumping of a moving electron to an adjacent empty well of the conveyor. This is justified if disorder in the device is sufficiently low. Experimental observations show that these types of errors are not detrimental to shuttling~\cite{Seidler2022,Struck2023,Xue2023} and if relevant they can be reduced by technological optimization.

\subsection{State Preparation and Measurement}

In the spin bus architecture, readout and initialization are performed in dedicated zones. In the modular architecture, it is assumed that one QD per register is coupled to a readout/initialization apparatus. The starting point of the initialization is the singlet ground state of two electrons in an auxiliary QD, which is separated into two QDs by an adiabatic sweep. For readout, the same adiabatic sweep is performed in reverse, merging the electron to be measured into one dot with a reference electron. Due to the Pauli exclusion principle only totally antisymmetric two-electron spin states are allowed in a single QD. Thus, depending on the regime of operation this either implements a singlet-triplet measurement or a spin-parity measurement~\cite{RevModPhys.95.025003}.

The adiabatic sweep can be performed with a fidelity as high as ${F_m = 0.999}$~\cite{ArqueProposal} and the subsequent charge detection verifying the outcome of the sweep can be expected to have an error probability of ${\mathcal O (10^{-5})}$ in devices optimized for spin shuttling~\cite{Seidler2022}. The high fidelity of the adiabatic process requires a relatively long time, however. We assume a readout time of ${T_r \approx \SI{5}{\micro\second}}$ per electron~\cite{ArqueProposal}.

\section{Performance Estimates in the Presence of Noise}

In this section, we evaluate the gate sequences devised in Sec.~\ref{sec_architectures}, including the errors described in Sec.~\ref{sec_error_model}. Subsequently, in Sec.~\ref{sec_transmons}, the result is put into context with other architectures and a comparison with superconducting transmon qubits is given.

\subsection{Performance of the Spin Qubit Architectures\label{sec_error_evaluation}}

The time evolution of the qubit density matrix $\rho$ is modeled by means of the Kraus operators for each operation. For simplicity, only one unit cell with periodic boundary conditions is modeled, undergoing one round of QAOA. This procedure will not return the correct output state of the algorithm, but is sufficient for estimating the physical errors by comparing the imperfect output state $\rho_\mathrm{err}$ with the ideal output state $\rho_\mathrm{id}$ without errors. Note that a single round of QAOA can deliver only a coarse approximation of the ground state and generally the performance of QAOA improves with increasing number of rounds~\cite{doi:10.1073/pnas.2006373117,Harrigan2021,doi:10.1126/science.abo6587,Weidenfeller2022scalingofquantum}. The fidelity 
\begin{equation}
F = \left(\mathrm{tr}\sqrt{\sqrt{\rho_\mathrm{id}} \rho_\mathrm{err} \sqrt{\rho_\mathrm{id}}}\right)^2\in [0, 1]
\end{equation}
is a measure for the probability to find the unit cell in its desired output state, where no qubit suffered from an error~\cite{NielsenChuang}. Thus, we introduce the average single qubit error probability $p_{1q}$. Assuming that errors on the four individual qubits are independent, $p_{1q}$ is obtained from ${F = (1 - p_{1q})^4}$.

\begin{figure*}
    \centering
    \includegraphics[width=0.99\textwidth]{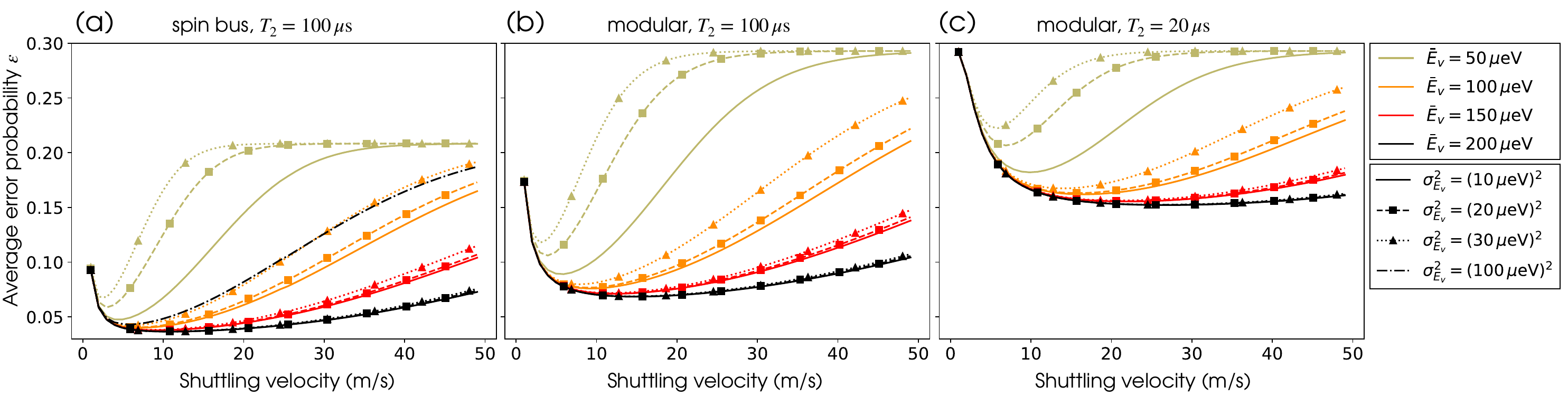}
    \caption{Average error probability after one round of QAOA for (a) the spin bus with ${T_2 = \SI{100}{\micro\second}}$ and (b) for the modular architecture with an optimistic $T_2 = \SI{100}{\micro\second}$ and (c) with a realistic ${T_2 = \SI{20}{\micro\second}}$. In all panels, the color (style) of each line encodes the mean $\bar E_v$ (variance $\sigma_{E_v}^2$) of the distribution of the valley splitting. The less weight the distribution has near ${E_v = 0}$, the better the algorithm performs and the broader the window of feasible shuttling velocities becomes. Since ${T_2 \ll T_1}$, the curves converge towards the scenario of a dephased qubit with almost intact spin projection.}
    \label{fig_error_rate}
\end{figure*}

The single round of QAOA is followed by a readout of all qubits. The probability to observe an error on a single qubit at the end of the total circuit is thus 
\begin{equation}
\varepsilon  = 1 - (1 - p_{1q}) F_r F_m, 
\end{equation}
where $F_r$ is the fidelity of the shuttling and idling in the readout step and $F_m$ is the fidelity of the measurement itself. This total physical single qubit error probability is plotted for the spin bus architecture in Fig.~\ref{fig_error_rate}(a) and for the modular architecture with both an optimistic assumption for $T_2$ that can possibly be achieved by dynamical decoupling in Fig.~\ref{fig_error_rate}(b) as well as with a realistic $T_2$ in Fig.~\ref{fig_error_rate}(c).

The results here are obtained with ${p_{\phi,\mathrm{idl}} = t/T_2}$.  In the case of a Gaussian decay, ${p_{\phi,\mathrm{idl}} = (t/T_2)^2}$ of the coherences due to low-frequency noise, the performance is found to be considerably better, in particular for slow shuttling. We discuss this case in Appendix~\ref{app_low_frequency_noise}. Realistically, a result between those extremes can be expected.

Both architectures show the same general dependence on the distribution of the valley splitting and the shuttling velocity, although the spin bus performs slightly better even in case of identical dephasing [cf. Fig.~\ref{fig_error_rate}(a) and (b)], despite the comparable amount of shuttling and idling. This is due to the fact that in the modular architecture a total of ten SWAP gates are required as additional steps, which are decomposed into 30 CP and 60 Hadamard gates, introducing additional errors, as discussed in Sec.~\ref{sec_architectures}.

The error probability $\varepsilon$ shows the interplay of the main dephasing mechanisms: If the algorithm requires a long time the qubits will strongly dephase due to their finite $T_2$. This is mitigated by increasing the shuttling velocity $v$ which lowers the error probability at first. However, as $v$ increases, the non-adiabatic errors of the shuttling increase, such that all curves finally converge to the case of a fully decohered spin for large $v$. This leads to the emergence of an optimal shuttling velocity. 

The probability distribution function of the valley splitting, characterized here by its mean $\bar E_v$ and variance $\sigma_{E_v}^2$, determines the strength of the non-adiabatic effects and thus the optimum. In particular, a distribution with a large weight near ${E_v = 0}$ is problematic for shuttling. Reducing the spread $\sigma_{E_v}^2$ of the distribution and increasing its mean $\bar E_v$ by engineering the interface of the semiconductor heterostructure will result in both a lower average error and a broader window of $v$ in which near-optimal results can be expected. Increasing the mean valley splitting to ${\bar E_v \approx \SI{200}{\mu\electronvolt}}$ with a standard deviation of ${\sigma_{E_v} \approx \SI{30}{\mu\electronvolt}}$ or less --- which is well within the theoretically predicted range of distributions~\cite{Lima_2023_2} --- a single qubit error probability as low as ${\varepsilon \approx 0.037}$ ({$\varepsilon \approx 0.069$}) is observed for the spin bus (modular) architecture. If the noise is dominated by low-frequency components, even lower error probabilities are expected (see App.~\ref{app_low_frequency_noise}).

Note that these estimates rely on averaged quantities. In an actual device, the errors may be far stronger than estimated here due to individual components deviating from the expected performance, e.g., a shuttling lane with a local dip of the valley splitting. Such a component can be avoided by adapting the shuttling sequence in order to minimize its harmful effects at the cost of detours and additional idling~\cite{ender_parity_2023}.

\subsection{Context and Comparison With Superconducting Qubits\label{sec_transmons}}

We finally assess the utility of the spin qubit device for near-term algorithms. Proposition 2 of Ref.~\cite{StilckFranca2021} states that there is a maximum depth for a quantum optimization algorithm that consists of a fraction $f_1$ ($f_2$) of single- (two-) qubit gate layers with local error probability $p_1$ ($p_2$): Above a total of
\begin{equation}
D_\mathrm{max} = \frac{\log \epsilon^{-1}}{2(f_1 p_1 + f_2 p_2)}
\end{equation}
layers, there is a classical algorithm which can sample from a Gibbs state in polynomial time, with an error ${\epsilon |\!| H_\mathrm{op} |\!|}$ with respect to the output of the noisy quantum algorithm. Assuming that the errors are dominated by shuttling and idling, rather than gate errors --- which is justified by the observations in Fig.~\ref{fig_error_rate} --- we estimate $p_{1(2)}$ by assuming that the errors are equally distributed over the shuttling and idling time. Then we weight them by the share of single and two-qubit layers of the total distance and duration, as discussed in Sec.~\ref{sec_spin_bus}.

For the spin bus architecture with optimal shuttling velocity we find that with ${\epsilon = 0.1}$ there are ${D_\mathrm{max} \gtrsim 280}$ gate layers possible before the quantum advantage is lost. While this estimated circuit depth is comparable with or less than what other hardware platforms with stationary qubits currently promise~\cite{StilckFranca2021,doi:10.1073/pnas.2006373117,Harrigan2021,doi:10.1126/science.abo6587}, we emphasize that this corresponds to ${p \approx 31}$ rounds of Parity QAOA independent of the system size. In more conventional architectures, where layers of SWAP gates are used to emulate the required connectivity between stationary qubits, the number of layers scales with the system size, thus with ${\mathcal{O}(10^2)}$ qubits $D_\mathrm{max}$ allows only a few rounds of QAOA~\cite{Weidenfeller2022scalingofquantum}. This astounding result is achieved by abstracting the problem to the Parity Architecture which naturally fits the spin bus topology. In the following we discuss further options for quantum error mitigation in order to recover the noisy output state.

For a more substantive comparison we, adapt the performance analysis to superconducting transmon qubits. For technical convenience, we assume a chip layout which matches the topology of the modular spin qubit architecture, where the interaction between the qubits is mediated by capacitive coupling, as opposed to nearest-neighbor exchange and spin shuttling. The resulting two-dimensional grid is composed of square and octagonal tiles alternately. The capacitive coupling allows for native CZ gates. The implementation of Parity QAOA for transmon qubits is analogous to the circuit presented in Sec.~\ref{sec_modular}, omitting the shuttling steps, which makes a comparison between the two hardware platforms straightforward. Note that we expect a square lattice of transmon qubits to perform slightly better due to the additional SWAP gates required by the modular architecture, as discussed in Sec.~\ref{sec_modular}.

The gate fidelity for transmon qubits can reach up to ${\approx 0.9999}$ for single-qubit gates and ${0.998-0.999}$ for two-qubit gates~\cite{transmon_sqg,sc_qubits_review}, although the in-system performance of simultaneous two-qubit gates is typically lower and can be $\approx 0.996\%$~\cite{transmon_google,doi:10.1126/science.adh9932}. We model this with depolarization channels for both single- and two-qubit gates with error probabilities of ${p_{d,1\mathrm{qg}} = 3\times 10^{-4}}$ and ${p_{d,2\mathrm{qg}} = 1\times 10^{-3} - 5\times 10^{-3}}$, respectively. These gates can be performed within ${T_{1q} \approx \SI{20}{\nano\second}}$ and ${T_{2q} \approx \SI{50}{\nano\second}}$. Realistic values for decoherence and relaxation times in current transmon devices are ${T_2\approx \SI{100}{\micro\second}}$ and $T_1 \approx \SI{115}{\micro\second}$~\cite{sc_qubits_review,Jurcevic_2021,transmon_dissipation}. We assume a readout and initialization fidelity of ${F_m = 0.995}$ and we assume that all qubits can be read out simultaneously~\cite{Jurcevic_2021}. Compared to the spin qubits, the gates of the transmons are faster and have similar gate fidelity (slightly lower for in-system performance) and with a similar $T_2$, and no shuttling is required.

We estimate the performance of the transmon chip in analogy to the spin qubits. With optimistic assumptions for the two-qubit gate fidelity we find that the error can be between ${\varepsilon \approx 0.043}$ for a choice of ${p_{d,2q} = 2\times 10^{-3}}$ and ${\varepsilon \approx 0.027}$ for a choice of ${p_{d,2q} = 10^{-3}}$, corresponding to ${F_\mathrm{CZ} \approx 0.9987}$ and ${F_\mathrm{CZ} \approx 0.9993}$, respectively. This result is slightly better than the optimal outcome for modular spin qubits, although of a comparable magnitude. We note that the optimal result obtained from the spin bus, whose topology naturally matches the Parity Architecture, lies well within the range defined by the optimistic transmon chip. 

Using the more conservative estimate of ${p_{d,2q} = 5\times 10^{-3}}$, corresponding to ${F_\mathrm{CZ} \approx 0.9967}$ which was observed for simultanoues two-qubit gates integrated in a two-dimensional grid of 67 qubits~\cite{transmon_google,doi:10.1126/science.adh9932}, we find ${\varepsilon \approx 0.087}$. Both spin qubit architectures can reach and exceed this performance. The spin bus with optimized shuttling velocity can outperform this transmon chip for all considered distributions of the valley splitting, leaving a wide margin for the optimization of the semiconductor hetereostructure.

In the limit of low-frequency noise resulting in a Gaussian decay of the coherences, Appendix~\ref{app_low_frequency_noise}, both spin qubit architectures can outperform the transmon qubits even for the optimistic choice of gate fidelities. The superconducting qubit platform is hardly affected by the choice of exponential or Gaussian decay, since the execution time of the algorithm is much faster than $T_2$, with gate errors being the limiting factor.

\section{Decoding and Error Mitigation\label{sec_decoding}}

In the previous section we assessed the expected single qubit error probabilities for one round of QAOA both with the spin bus and a modular architecture. To evaluate the total algorithm performance, one needs to consider that parity QAOA is typically evaluated by measuring all qubits and then reconstructing the logical state from so-called spanning trees---subgraphs of ${N-1}$ physical qubits connected by exactly one path that spans over all $N$ logical qubits~\cite{PhysRevA.108.032408}. One such spanning tree is sufficient for reconstructing the logical state up to a global spin-flip. In the error-free case all spanning trees yield the same logical state, which corresponds to the output state of the algorithm. If, however, physical errors occur, different logical states are obtained from different spanning trees. In that case, the logical state with the lowest energy is accepted as the optimal result. Considering the error probabilities computed in the previous section, both architectures operated with optimal shuttling velocity can reach the low-error regime where Parity QAOA has a clear advantage over conventional QAOA: Simulations of small noisy systems have shown that Parity QAOA equipped with the classical post-processing of the spanning trees can still have a high success probability, even exceeding the success probability of standard QAOA with an ideal system~\cite{PhysRevA.108.032408}.

This is sufficient for the treatment of optimization problems, although it does not allow for a decision on whether the accepted result is the output of the algorithm or was produced by random noise. For some instances, it may also be beneficial to decode the readout results in a way that allows the reconstruction of the output state, which is particularly crucial in view of future applications of universal Parity quantum computing~\cite{PhysRevLett.129.180503}. We continue to evaluate the performance with respect to those two aspects. In Ref.~\cite{PhysRevA.93.052325}, an estimate is given for an upper bound for the probability of the decoding to fail and result in the acceptance of the wrong logical state from the Parity Architecture.

This upper bound for the decoding error probability decays exponentially with the number of logical qubits. Based on the estimated wiring complexity, we assume that a spin bus processor with up to 50 unit cells can be operated with room temperature controls~\cite{ArqueProposal}, which allows for up to 20 logical qubits. Together with the physical error probability ${\varepsilon \approx 0.037}$ (${\varepsilon \approx 0.069}$) of the spin bus (modular) architecture, this results in a decoding error probability of $\leq 2.5\%$ ($\leq 13.2\%$) with realistic gate errors. This represents an upper bound only and the actual probability is expected to be much lower with a sophisticated decoding scheme. For example, when applying belief propagation, the decoding error probability can be expected to be below $1\%$ for 6 (8) or more logical qubits, i.e., 15 (28) physical qubits~\cite{PhysRevA.93.052325}.

For the spanning tree readout we expect that, with a reliable algorithm, the correct output state is observed more frequently than random states. This can be exploited to decide whether the accepted result is also the correct output. To do so, we assume that $n$ spanning trees of ${N-1}$ qubits are read out and used for decoding. We also assume that they are distributed evenly over the chip in order to access all information, such that all qubits are included in one tree before any qubit is included in a second. With $N$ logical and thus ${K = N(N-1)/2}$ physical qubits, this means that all physical qubits are part of ${\lfloor n(N-1) / K \rfloor = \lfloor 2n/N \rfloor}$ or ${\lfloor 2n/N \rfloor+1}$ spanning trees. The number of qubits that are part of ${\lfloor 2n/N \rfloor+1}$ trees is ${n(N-1) \mod K}$.

Thus, with one physical error on the chip, there is a chance of finding ${\lfloor 2n/N\rfloor}$ incorrect trees with probability ${1 - [n(N-1) \mod K] / K}$ and of finding ${\lfloor 2n/N\rfloor + 1}$ incorrect trees with probability ${[n(N-1) \mod K] / K}$. The expected number of incorrect spanning trees with one physical error is thus
\begin{equation}
    \langle n_\mathrm{inc} \rangle(1) =  \left\lfloor \frac{2n}{N} \right\rfloor  +  \frac{1}{K}\, \left[n(N-1) \mod K\right].
\end{equation}

Each further error will also introduce $\langle n_\mathrm{inc} \rangle(1)$ incorrect spanning trees, but has an increasing chance to affect trees that already return the incorrect logical state. Thus, the expected number of incorrect results can be computed recursively for the $m$th error
\begin{eqnarray}
    \langle n_\mathrm{inc} \rangle(m) &=& \langle n_\mathrm{inc} \rangle(m-1) \label{eq_n_inc_m}\\
    && + \max\left[1- \frac{\langle n_\mathrm{inc} \rangle(m-1)}{n}, 0\right] \langle n_\mathrm{inc} \rangle(1),\nonumber
\end{eqnarray}
where the $\max$ was included in order to avoid artifacts from the discrete computation. Consequently, with $m$ physical errors, on average 
\begin{equation}
\langle n_\mathrm{ok} \rangle(m) = n - \langle n_\mathrm{inc} \rangle(m)
\end{equation}
spanning trees can be expected to return the correct logical state.

Inverting $\langle n_\mathrm{ok} \rangle(m)$ and assuming that the physical errors are independent and equally distributed allows us to find the probability distribution
\begin{equation}
P_{\varepsilon, K}(\langle n_\mathrm{ok} \rangle) = B[m(\langle n_\mathrm{ok} \rangle) | \varepsilon, K]
\end{equation}
of $\langle n_\mathrm{ok}\rangle$, where $B(m | \varepsilon, K)$ is a binomial distribution of the number of errors $m$ with the physical error probability $\varepsilon$ and qubit number $K$. We now define $\langle n_\mathrm{ok}\rangle_x$ such that in a fraction $x$ of the experimental runs more than $\langle n_\mathrm{ok}\rangle_x$ spanning trees returning the correct output state can be expected:
\begin{equation}
    \sum_{\langle n_\mathrm{ok} \rangle = 0}^{\langle n_\mathrm{ok} \rangle_x} P_{\varepsilon, K}(\langle n_\mathrm{ok} \rangle) = 1-x.
\end{equation}
A logical state is accepted once it is obtained from more than $\langle n_\mathrm{ok} \rangle_x$ spanning trees, where $x$ parameterizes the expectation that a fraction $x$ of the runs should allow for a positive decision to be made. Choosing a smaller $x$ will result in a lower rate of erroneous positive decisions at the cost of requiring more runs.

As a measure of the reliability of the algorithm, we compute
the probability for obtaining at least $\langle n_\mathrm{ok}\rangle_x$ spanning trees returning an identical logical state from an entirely random outcome of the algorithm ($\varepsilon = 0.5$): %is
\begin{equation}
    p_\mathrm{fail} = \sum_{\langle n_\mathrm{ok}\rangle = \langle n_\mathrm{ok} \rangle_x }^{n} P_{0.5, K}(\langle n_\mathrm{ok}\rangle).
\end{equation}
This estimate can be viewed as a statistical test based on the probability distribution $P_{\varepsilon, K}$, whether the output state could plausibly arise from a random result.

\begin{figure}
    \centering
    \includegraphics[width=0.49\textwidth]{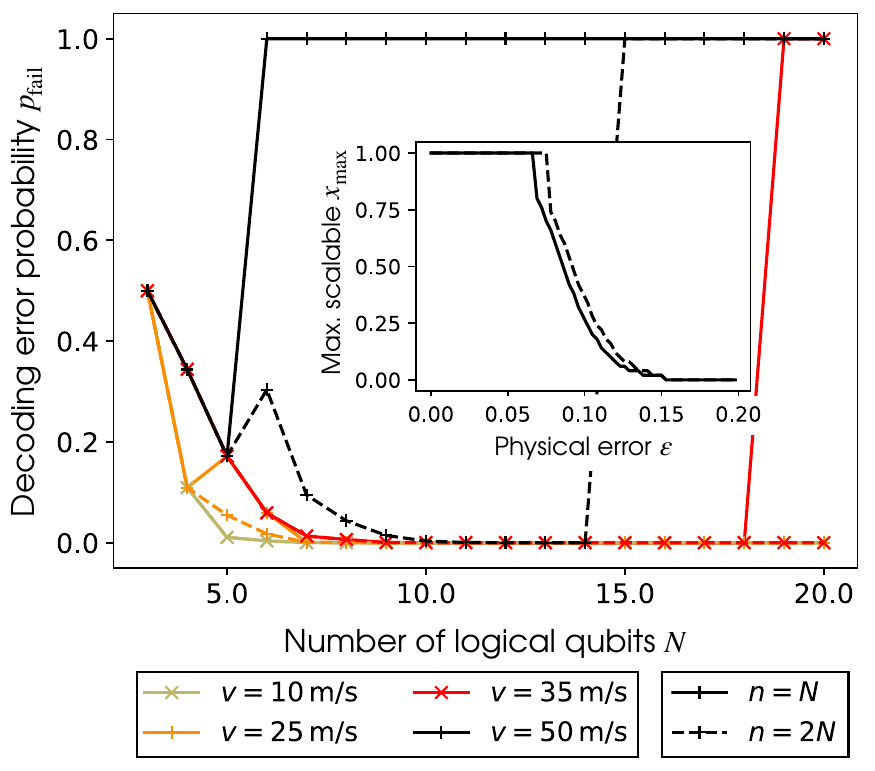}
    \caption{Decoding based on the readout of spanning trees for up to 20 logical qubits. (Main panel) Probability $p_\mathrm{fail}$ of accepting a random state as algorithm output computed with ${x=0.9}$ for the spin bus as a function of the number of logical qubits, with $\bar E_v = \SI{100}{\mu\electronvolt}$, $\sigma_{E_v}^2 = (\SI{20}{\mu\electronvolt})^2$. A number of ${n = N}$ ($2N$) spanning trees are decoded in the solid (dashed) curves. Increasing the system typically suppresses the decoding error, however, small jumps due to the discrete nature of spanning trees may occur (orange curve) and for a too large physical error the decoding may not be scalable. The requirements can be lowered by increasing the number $n$ of spanning trees. The lines only serve as a guide to the eye.
    (Inset) Maximal fraction $x$ of experiments, $x_\mathrm{max}$, that allows a decision about which state is the correct output state for any system size (scalable decoding) as a function of the error probability $\varepsilon$. If ${x = 1}$ the output state is reliably recovered from the decoding of the spanning trees for all qubit numbers, if ${x = 0}$ no decision is possible on whether the output is random or not, from a certain system size upwards. In between, decoding is still possible in general but works only for a %small 
    fraction $x$ of all attempts.}
    \label{fig_decoding}
\end{figure}

The probability $p_\mathrm{fail}$ of accepting a random state as algorithm output, regardless of it being the best solution to the problem or not, is shown in Fig.~\ref{fig_decoding} for the example of the spin bus architecture with different shuttling velocities, ${x = 0.9}$ and the choices of ${n=N}$ and ${n=2N}$. Naturally, $p_\mathrm{fail}$ decreases with the qubit number because the probability for repeatedly observing a random result decreases fast with system size. The curves may show a small jump due to the discrete nature of the spanning trees, since a change in $\langle n_\mathrm{ok} \rangle_{0.9}$ by 1 can be significant for a low number of logical qubits, this can be seen in the solid orange and dashed black curves.

Some curves with a high $\varepsilon$ furthermore show a big jump to $p_\mathrm{fail} = 1$. This can be explained by the fact that, for a too large error $\varepsilon$ and a too strict $x$, $\langle n_\mathrm{ok} \rangle_x$ will also decrease when the number of qubits is increased and may fall to 0 such that no decoding is possible. In that case, the decoding is not scalable. By increasing the number of spanning trees, $n$, more redundant information is used and thus higher error rates can be tolerated.

In the inset of Fig.~\ref{fig_decoding}, $x_\mathrm{max}$ is shown, which is the maximal $x$ where $\langle n_\mathrm{ok} \rangle_x$ does not decrease with the qubit number, as a function of the physical error probability. For ${x_\mathrm{max} = 1}$ it is possible to reliably recover the output state of the algorithm and increasing the number of qubits improves decoding. For ${x_\mathrm{max} = 0}$, the decoding may work for small systems but will fail at a certain qubit number, that is, the algorithm will still produce a candidate solution of the optimization problem but it is impossible to decide whether the result was produced by the algorithm or noise. In the intermediate regime, the decoding still works as usual and increasing the system size improves the error mitigation capabilities, but also introduces a finite probability that no decision can be made on the output of the algorithm. Based on the results of Sec.~\ref{sec_error_evaluation}, Parity QAOA can be performed and decoded reliably with the spin bus and the modular architecture with an optimistic value for $T_2$, achieved by dynamical decoupling.

\section{Summary and Conclusions\label{sec_conclusions}}

In this paper we investigated the implementation and performance of the Parity QAOA algorithm on two different electron spin qubit architectures, one based on electrons sparsely distributed over intersecting shuttling lanes (spin bus) and one where ${2\times 2}$ arrays of QDs form a lattice of registers interconnected by shuttling. We presented shuttling and gate sequences for realizing all elements of Parity QAOA on both platforms. While straightforward for the spin bus, the modular chip layout discussed here requires 10 successive SWAP gates to achieve the required connectivity. Alternatively, SWAP gates can be traded for hopping between adjacent QDs.

In order to consider realistic errors, we developed a model which allows for estimating the mean shuttling error as a function of the probability distribution of the valley splitting and the valley phase. Assuming Si/SiGe as host material and conveyor mode shuttling with realistic parameters for gate error rates, dephasing and valley splitting, we find that both architectures can complete one round of Parity QAOA with a low single-qubit error probability. The spin bus slightly outperforms the modular architecture due to the need for additional SWAP gates in the latter. The performance delivered by both platforms for Parity QAOA after optimizing the shuttling process can exceed the results expected from typical superconducting transmon qubits. This result not only points out the synergy between the Parity Architecture and spin qubit hardware but also shows that spin qubits can be serious contenders among the leading quantum computing platforms.

Our error analysis suggests that the main limitations are dephasing from inter-valley transitions if the shuttling is non-adiabatic and dephasing from environmental noise if the shuttling is too slow. Engineering the Si/SiGe interface~\cite{PaqueletWuetz2022,Lima_2023_1,Lima_2023_2} and Si quantum well~\cite{PhysRevB.104.085406,McJunkin2022} to deterministically enhance the valley splitting or adjusting the shuttling path and velocity to avoid excitations can
%seems the most viable way to 
further reduce the minimal error by allowing for a higher shuttling velocity. Protecting the spins from charge noise and prolonging their lifetime by dynamical decoupling can make the algorithm more robust at low velocities as well.

Finally, we discussed the possibilities for quantum error mitigation based on the results of the error model. While Parity QAOA has an intrinsic error mitigation capability, we show that it is not guaranteed that the actual outcome of the algorithm can be identified. Nevertheless, our estimations indicate that the physical errors of both spin qubit platforms are low enough to decode the final state with a high success probability. This is accomplished either with dedicated decoding schemes such as belief propagation or with the classical post-processing of spanning trees that is commonly used in conjunction with Parity QAOA.

Other studies have shown that a direct implementation of QAOA requires much higher gate fidelities than currently available in any platform to achieve a quantum advantage for problems whose coupling topology does not match that of the hardware~~\cite{Weidenfeller2022scalingofquantum,doi:10.1073/pnas.2006373117,Harrigan2021,doi:10.1126/science.abo6587}. The combination of the parity encoding with an architecture that is well-matched to its requirement paints a more optimistic picture. The cost for this advantage is a quadratic overhead in the number of qubits. Importantly, this increase in qubit number can be leveraged without requiring a higher fidelity, implying a much better scalability of parity encoded QAOA problems.

Our results highlight that a two-dimensional spin qubit platform can indeed serve as a natural implementation of the Parity Architecture even if the connectivity does not directly correspond to a square-lattice geometry. Thus, the utility of spin qubits for quantum computing tasks such as solving optimization problems can be advanced by alleviating the need for long-range interaction and by allowing for constant-depth QAOA with the possibility of quantum error mitigation. The results presented here show that Parity QAOA may be in reach of near-term spin qubit devices, promising a substantial quantum advantage for a large class of problems once a qubit number on the order of a few thousand is achieved. It can be expected that future improvements of the qubit coherence make universal Parity quantum computing viable. This will provide an advantage by reducing the circuit depth of cornerstone quantum algorithms such as the quantum Fourier transform, while relying exclusively on nearest-neighbor interactions and single-qubit gates~\cite{PhysRevLett.129.180503}.

We note that we considered only a minimal modular architecture, the performance with larger registers is an open question to be addressed in the future. Other promising directions are the use of resonator~\cite{Mi2028,PhysRevApplied.15.044052,PhysRevB.99.245306,PhysRevB.100.245427,PhysRevB.108.125437} and RF readout~\cite{PhysRevLett.110.046805,PhysRevX.8.041032,West2019,Urdampilleta2019} of spin qubits for measurement-based Parity quantum computing~\cite{Messinger2023} and the use of a hybrid formulation of Parity QAOA which reduces the number of constraints that are enforced explicitly and allows a modularization of the code~\cite{PRXQuantum.3.030304}. The latter can be useful for the evolution of the modular design with larger registers. For future research in order to unlock the full potential of spin qubits equipped with the Parity Architecture it is also relevant to explore whether electron or hole spin qubits can provide special advantages and it may be beneficial to revisit and further optimize the native gate set of the platform for the interactions required here.

%%%%%%%%%%%%%%%%%%%%%%%%%%%%%%%%%%%%%%%%%%%%%%%%%%%%%%%%%%%%%%%%%
%%%%%%%%%%%%%%%%%%%%%%%% Acknowledgments %%%%%%%%%%%%%%%%%%%%%%%%%%%%%
%%%%%%%%%%%%%%%%%%%%%%%%%%%%%%%%%%%%%%%%%%%%%%%%%%%%%%%%%%%%%%%%%

\begin{acknowledgments}
We thank Philipp Aumann, Adu Offei-Danso, Javad Kazemi, Enrique Naranjo Bejarano and Anette Messinger for helpful discussions, and Kilian Ender for critical comments on the manuscript.
This work was supported by the Austrian Research Promotion Agency (FFG Project No. P7050-026-023). This research was funded in whole, or in part, by the Austrian Science Fund (FWF) SFB BeyondC Project No. F7108-N38, through a START grant under Project No. Y1067-N27 and I 6011, by the
German Research Foundation (DFG) under Germany’s Excellence Strategy - Cluster of Excellence Matter and Light for Quantum Computing (ML4Q) EXC 2004/1-390534769 and by the Federal Ministry of Education and Research (Germany), Funding reference number: 13N15652.
For the purpose of open access, the author has applied a CC BY public copyright license to any Author Accepted Manuscript version arising from this submission. This project was funded via Project Si-QuBus within the QuantERA ERA-NET Cofund in Quantum Technologies and
within the QuantERA II Programme that have received funding from the European Union's Horizon 2020 research and innovation program under Grant Agreement No. 101017733. 
\end{acknowledgments}

%%%%%%%%%%%%%%%%%%%%%%%%%%%%%%%%%%%%%%%%%%%%%%%%%%%%%%%%%%%%%%%%%
%%%%%%%%%%%%%%%%%%%%%%%%% Appendices %%%%%%%%%%%%%%%%%%%%%%%%%%%%%%
%%%%%%%%%%%%%%%%%%%%%%%%%%%%%%%%%%%%%%%%%%%%%%%%%%%%%%%%%%%%%%%%%

\appendix
\section{Alternative Sequence for the Modular Architecture\label{app_AltMod}}

\begin{figure}[htb]
    \centering
    \includegraphics[width=0.49\textwidth]{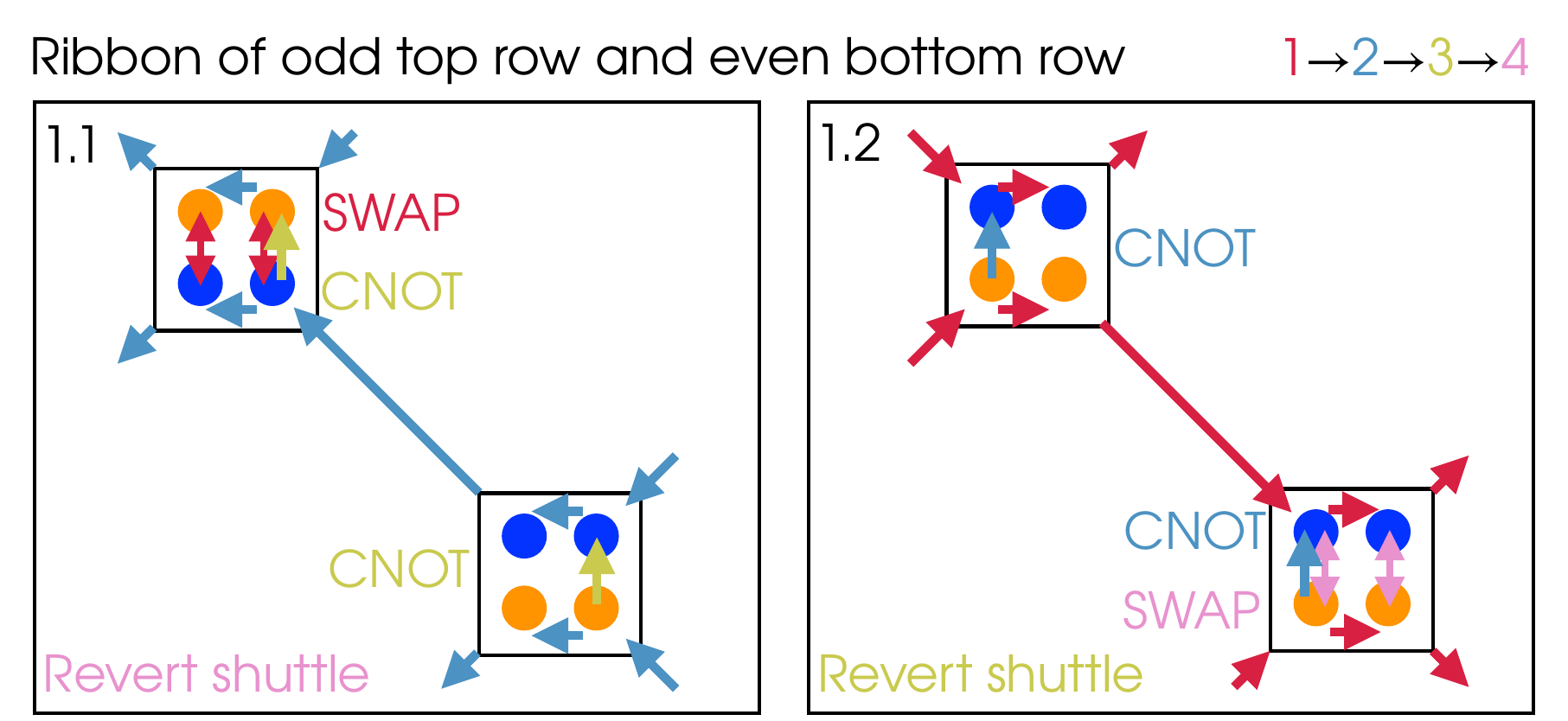}
    \caption{Alternative circuit for implementing the parity constraints on ribbons of an odd top and an even bottom row in a modular architecture if executed on each unit cell in parallel. This sequence can replace the steps 1.1 and 1.2 of Fig.~\ref{fig_modular_odd-even} and its reverse can replace step 4 of Fig.~\ref{fig_modular_odd-even}. Effectively, some SWAP gates are replaced by transitions between the QDs of a module, indicated by horizontal arrows. Thus, four SWAPs in steps 1 and 4 each can be avoided. The remaining SWAP gates are required for the subsequent steps.}
    \label{fig_modular_alternative}
\end{figure}

\begin{figure*}[htb]
    \centering
    \includegraphics[width=0.99\textwidth]{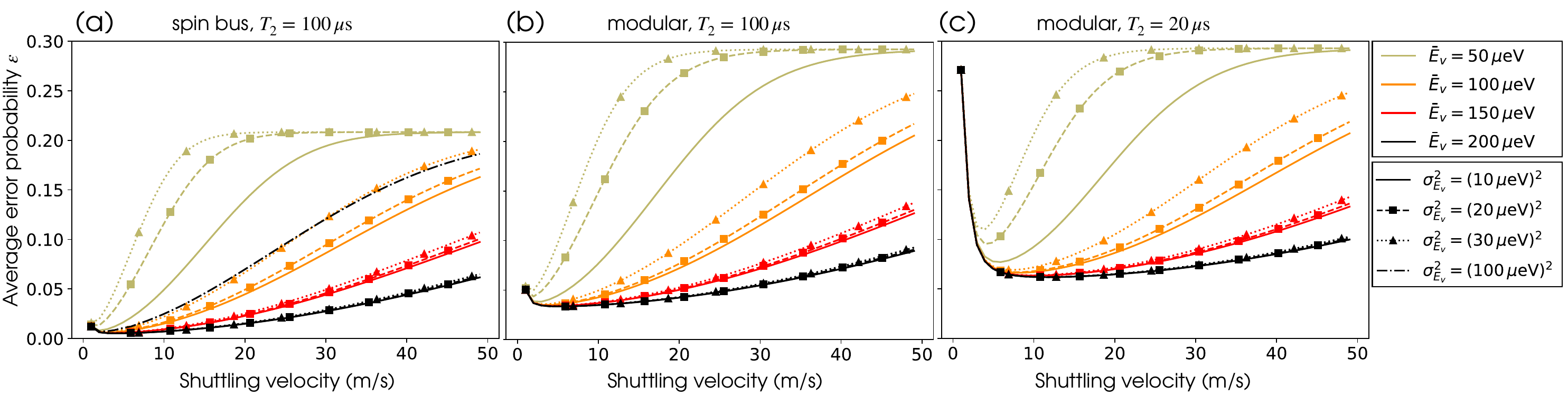}
    \caption{Average error probability after one round of QAOA with ${p_{\phi,\mathrm{idl}} = (t/T_2)^2}$ for (a) the spin bus with $T_2 = \SI{100}{\micro\second}$ and (b) for the modular architecture with an optimistic ${T_2 = \SI{100}{\micro\second}}$ and (c) with a realistic ${T_2 = \SI{20}{\micro\second}}$. The examples plotted here correspond to Fig.~\ref{fig_error_rate} in the case of quasistatic noise. The reduced dephasing during short idling times improves the performance for slow shuttling, while the dephasing due to non-adiabatic effects at high $v$ is unaffected.}
    \label{fig_error_Gaussian}
\end{figure*}

For the modular architecture, the parity constraints can also be realized with a circuit where eight SWAP gates are traded for coherent transitions between the QDs within a unit cell and thus requires only two SWAP gates per qubit. This alternative sequence is shown in Fig.~\ref{fig_modular_alternative}. The amount of shuttling operations is the same, with the exception that the electrons are now shuttled to vacant dots within the neighboring unit cells instead of adjacent sites, and the other gates remain unchanged. Thus, this approach can be a beneficial alternative if the fidelity of one intra-module transfer outperforms a SWAP gate.

With our assumptions, the fidelity of a single SWAP is expected to be ${\gtrsim 99.6\%}$. The transport within the module can be realized by utilizing the plunger and barrier gates for conveyor mode-shuttling or by phase-coherent bucket brigade shuttling in a double quantum dot. A fidelity per hop approaching this threshold was demonstrated in SiMOS devices~\cite{Yoneda2021,Noiri2022-2}. In Si/SiGe, the probability of spin-flip errors has been shown to be in the required range for high-fidelity shuttling~\cite{PRXQuantum.4.030303} and theoretical estimates predict a fidelity above the threshold to make this alternative viable~\cite{PhysRevB.102.195418}.

\section{Performance in the Presence of Low-Frequency Noise\label{app_low_frequency_noise}}

The dominating source of dephasing in semiconductor spin qubits is considered to be charge noise, electric field fluctuations with a power spectral density ${S(\omega) \propto 1/\omega^\alpha}$, ${\alpha\approx 1}$~\cite{RevModPhys.95.025003}. The low-frequency noise gives rise to a Gaussian decay of the coherences. Compared to the case of rapid fluctuations, this leads to an improved performance of the shallow algorithm discussed here for instances where the passive dephasing is the limiting factor. We investigate the case of quasistatic noise by estimating the average single-qubit error probability $\varepsilon$ with ${p_{\phi,\mathrm{idl}} = (t/T_2)^2}$ as probability for the dephasing channel during the idling of the qubits. The results are depicted in Fig.~\ref{fig_error_Gaussian}.

As a consequence, the error probability for a low shuttling velocity $v$ is considerably reduced. Thus, the optimal result is found for slower shuttling and with a lower minimal error probability. Towards higher shuttling velocity, where the errors are dominated by non-adiabatic effects, the effect vanishes. With a mean valley splitting of ${\bar E_v \approx \SI{200}{\mu\electronvolt}}$ and a standard deviation of ${\sigma_{E_v} \approx \SI{30}{\mu\electronvolt}}$ or less a single qubit error around ${\varepsilon \approx 0.005}$ (${\varepsilon \approx 0.034}$) is observed for the spin bus (modular) architecture.

\FloatBarrier

%%%%%%%%%%%%%%%%%%%%%%%%%%%%%%%%%%%%%%%%%%%%%%%%%%%%%%%%%%%%%%%%%
%%%%%%%%%%%%%%%%%%%%%%%%% Literature %%%%%%%%%%%%%%%%%%%%%%%%%%%%%%
%%%%%%%%%%%%%%%%%%%%%%%%%%%%%%%%%%%%%%%%%%%%%%%%%%%%%%%%%%%%%%%%%

\bibliography{literature.bib}

\end{document}